**Contact feedback helps snake robots propel against uneven terrain using vertical bending**

Qiyuan Fu* and Chen Li

Department of Mechanical Engineering, Johns Hopkins University, Baltimore, MD 21218, USA

*Corresponding authors. E-mail: [redacted]

## Abstract

Snakes can bend their elongate bodies in various forms to traverse various environments. We understand well how snakes use lateral body bending to push against asperities on flat ground for propulsion, and snake robots can do so effectively. However, snakes can also use vertical bending to push against uneven terrain of large height variation for propulsion, and they can adjust this bending to adapt to novel terrain presumably using mechano-sensing feedback control. Although some snake robots can traverse uneven terrain, few have used vertical bending for propulsion, and how to control this process in novel environments is poorly understood. Here we systematically studied a snake robot with force sensors pushing against large bumps using vertical bending to understand the role of sensory feedback control. We compared a feedforward controller and four feedback controllers that use different sensory information and generate distinct bending patterns and body-terrain interaction. We challenged the robot with increasing backward load and novel terrain geometry that break its contact with the terrain. We further varied how much the feedback control modulated body bending to conform to or push against the terrain to test their effects. Feedforward propagation of vertical bending generated large propulsion when the bending shape matched terrain geometry. However, when perturbations caused loss of contact, the robot easily lost propulsion or had motor overload. Contact feedback control resolved these issues by helping the robot regain contact. Yet excessive conformation interrupted shape propagation and excessive pushing stalled motors frequently. Unlike that using lateral bending, for propulsion generation using vertical bending, body weight that can help maintain contact with the environment but may also overload motors. Our results will help snake robots better traverse uneven terrain with large height variation and can inform how snakes use sensory feedback to control vertical body bending for propulsion.

## 1. Introduction

With their slender and highly flexible body, snake robots hold the promise as a versatile platform to traverse diverse environments [1], especially complex 3-D terrain with large obstacles [2], [3] that challenge wheeled and legged robots. Similar to snakes [4]–[6], many snake robots have been developed to use lateral bending to push against vertical structures on the sides (hereafter referred to as lateral push points) to move on flat surfaces [7]–[12]. However, the real world is rarely flat but often three-dimensional. Snakes also traverse 3-D terrain with large height variation but lacking lateral push points, such as climbing over large boulders and fallen trees [13]–[15]. In contrast, snake robots in 3-D environments are still inferior to snakes in versatility and efficiency.

A main reason for this lack is that we do not yet well understand how to use vertical bending to generate propulsion by pushing against uneven terrain of large height variation below the body (hereafter referred to as vertical push points). On flat surfaces, some snake robots slightly lift body sections that experience large drag to improve efficiency [16], [17]. When traversing 3-D environments, many snake robots use actuated wheels or treads for propulsion [1]. Some snake robots form a rolling loop in the vertical plane to traverse small obstacles [18], [19], which can be unstable due to the narrow base of support. A few snake robots bend into shapes designed based on the geometry of dedicated terrain such as steps or pipes [20]–[24]. During their movement, vertical bending is only used to connect body sections performing distinct movement patterns, such as lateral undulation above and below a step [20], [22]–[24] or wrapping around different pipes [20], [21]. Several snake robots traverse uneven terrain using gaits designed for level ground, such as a sidewinding-like gait or lateral undulation, but suffer from severe slipping [21], [25], [26]. They accommodate the height variation by bending body vertically using controlled compliance but the vertical bending contributes little to propulsion. Only a few robots have deliberately used vertical bending to push against vertical push points for propulsion, but they cannot adapt to novel terrain without human operation [15], [27]–[30] or only move on ideally smooth surfaces without considering external forces such as friction and gravity [7], presumably because of lacking understanding of how to control vertical bending to adapt to novel environments.

Recent studies revealed that generalist snakes can traverse uneven terrain with large height variation by using vertical bending to push against vertical push points. The corn snake can traverse a row of horizontal cylinders by propagating a vertical wave posteriorly to push against the cylinders (figure 1(a)) [15]. The snake appears to coordinate contact forces in the vertical plane from multiple push points for propulsion or braking, as evidenced by the variable fore-aft force measured on one of the cylinders (figure 1(a), bottom). When in a narrow channel with a wedge, the corn snake initially uses a concertina gait (figure 1(b), red, wiggly body section) to brace against vertical walls to slowly (3.4 cm/s) move forward [15]. But once it gains substantial contact with the wedge, it transitions to propagating vertical bending posteriorly to push against the wedge (figure 1(b), blue) to move forward more rapidly (7 cm/s). The corn snake can also traverse a 3-D uneven terrain by propagating a 3-D bending posteriorly (figure 1(c)) [31]. During traversal, a similar number of potentially propulsive contact points are formed by the snake combining lateral and vertical bending. For example, a laterally bending body section (figure 1(c), blue bands) can potentially push against vertical edges of higher blocks lateral to the body (yellow squares), while a vertically bending body section (figure 1(c), red bands) can potentially push against horizontal edges of blocks below the body (blue squares) [31]. This combination that includes body bending in 3-D may provide snakes with substantially more push points to coordinate contact forces to improve stability, maneuverability, and efficiency in complex 3-D environments than using lateral bending alone, which the majority of previous terrestrial snake locomotion studies have focused on [32], [33].

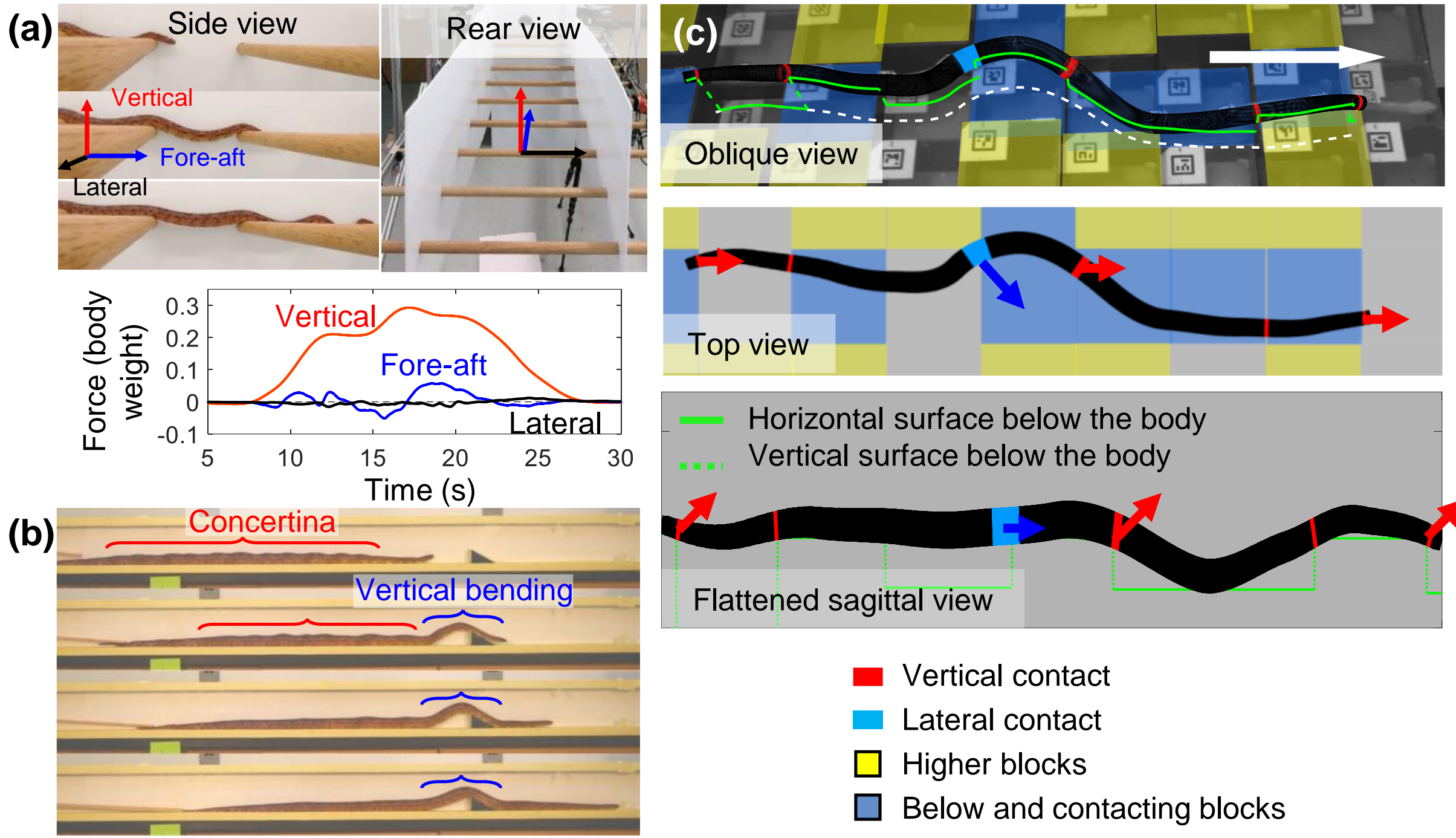

**Figure 1. Observations of snakes using vertical body bending. (a)** A snake traversing a row of horizontal cylinders by propagating a vertical wave posteriorly. Adapted from [15]. Top left: representative side view snapshots. Top right: rear view of setup. Bottom: Measured terrain reaction forces exerted on the snake by a cylinder along vertical (red), fore-aft (blue), and lateral (black) directions. **(b)** Side view of a snake traversing a wedge in a narrow tunnel (between the back wall and a transparent front wall) using vertical bending. Adapted from [15]. **(c)** Oblique, top, and flattened sagittal view of a snake traversing a 3-D uneven terrain by combining vertical and lateral bending. Adapted from [31]. Red and blue bands show vertically and laterally bending body sections that potentially push against the terrain, respectively. White arrow shows direction of movement.

This shape propagation is likely sensory-modulated, considering a generalist snake's ability to adjust its body bending patterns to adapt to novel terrain. For example, when encountering vertical structures with various spatial configurations on a flat surface, generalist snakes adjust lateral body bending patterns to maintain pushing against these lateral push points [5], [6]. Snakes likely use multiple senses to guide this modulation. When traversing uneven terrain with large height variation, the corn snake steers its head laterally and dorsoventrally more frequently, presumably using vision [34] to select a path [35], while the rest of its body simply follows its path [31]. In addition, snakes have tactile receptors within the skin to sense contact forces [36], stretch receptors within the muscles and tendons to sense body shape indirectly [37], and gravity sensation [38]. These internal or external mechanosensation may provide additional cues for the modulation of bending patterns, such as in a narrow tunnel on a horizontal plane where vision is unlikely to modulate bending after the head [4]. In addition, a snake can modulate the lateral bending pattern when the body slips out of the path of its head or to push harder against existing contact points [5], [6], presumably with contact feedback control.

Decoding control principles by measuring and manipulating neural activities is challenging in snakes because their complex sensory nervous systems are not well characterized [39]. To understand how to modulate body bending to effectively push against the environment for propulsion using sensory feedback, using snake robots as robophysical models is an amenable approach, because they allow controlled variation of control strategies and repeatable experiments [40]. Many robot studies have been used to investigate this modulation for a laterally bending snake, either by "passively" conforming to novel terrain geometry using controlled compliance or by actively exploiting sensed push points [2], [12]. However, it is unclear how a snake controls vertical bending to generate propulsion against the environment and whether and how sensory feedback control helps it adapt to perturbations. Understanding this question will not only inform why snakes use vertical body bending to traverse various 3-D terrain (figure 1), but also allow snake robots to exploit more terrain surfaces for propulsion generation to better traverse similar 3-D environments.

Here we used a snake robot instrumented with force sensors as a robophysical model to understand this question. We drove it to traverse a continuous track with large height variation using vertical bending. To understand whether and how sensory feedback control helps vertical bending to effectively produce propulsion, we compared a feedforward controller and four feedback controllers that modulate bending patterns in distinct ways. These controllers were designed to reproduce behaviors observed in snakes, such as propagation of body bending posteriorly [15], [41], exploration behavior of the head [15], [31], and modulation of bending patterns posterior to the head [5], [6], [41]. Considering the seemingly similar manner in pushing against suitably oriented terrain surfaces for propulsion by propagation of lateral bending [2], [12], [32] and vertical bending [15], [31], we adapted two types of contact feedback-controlled modulation that were hypothesized for laterally bending snakes previously: controlled conformation to the terrain [42] and active pushing against the terrain [12], [28]. We varied the degree of them to reveal their effects on vertical bending qualitatively. To understand why the distinct bending patterns generated by different controllers affected the performance, we analyzed how body-terrain contact was modulated by the bending patterns and affected the performance.

To test whether and how well each controller adapts to various perturbations, we tested the robot's success rates of traversal in five cases with various additional backward loads or novel terrain geometry. We hypothesized that the robot can traverse the uneven terrain by propagating a vertical bending shape posteriorly without using lateral bending. If contact is maintained, the propulsion generated can also increase to accommodate additional backward loads. We also hypothesized that the robot with feedforward control will struggle more when terrain geometry is changed. In contrast, contact feedback control can enable higher success rates than feedforward control.

Finally, we discussed the difference between propulsion generation using vertical bending with that using lateral bending.

## 2. Methods

### 2.1. Robophysical model

We used a snake robot (1.18 m long, 3.0 kg) with 9 pitch and 9 alternating yaw joints from our previous study [23] as the robophysical model (figure 2(a)). The alternating joint structure is common to produce 3-D motions similar to snakes [33]. Because we only studied vertical bending here, the yaw joints were fixed to be straight. All the pitch joints were actuated by servo motors (Dynamixel XM430-W350-R, ROBOTIS, Lake Forest, CA, USA). Each motor sent the present joint angle and motor current to a desktop computer and received the goal angle or current commands at a frequency of 31 Hz (supplementary material section 1.1). Each motor used an internal feedback controller to reach the goal angle or current it received [43]. Each motor disabled torque output automatically after being overloaded by a large external torque. Each robot link consisted of one pitch joint (figure 2(a), cyan) motor and one yaw joint (orange) motor. Because the most posterior pitch joint was reserved for lifting an active wheel (supplementary material section 1.2) used to propel the robot to the initial position before each locomotion experimental trial (section 2.3), we excluded it from the vertical bending control and the analyses hereafter.

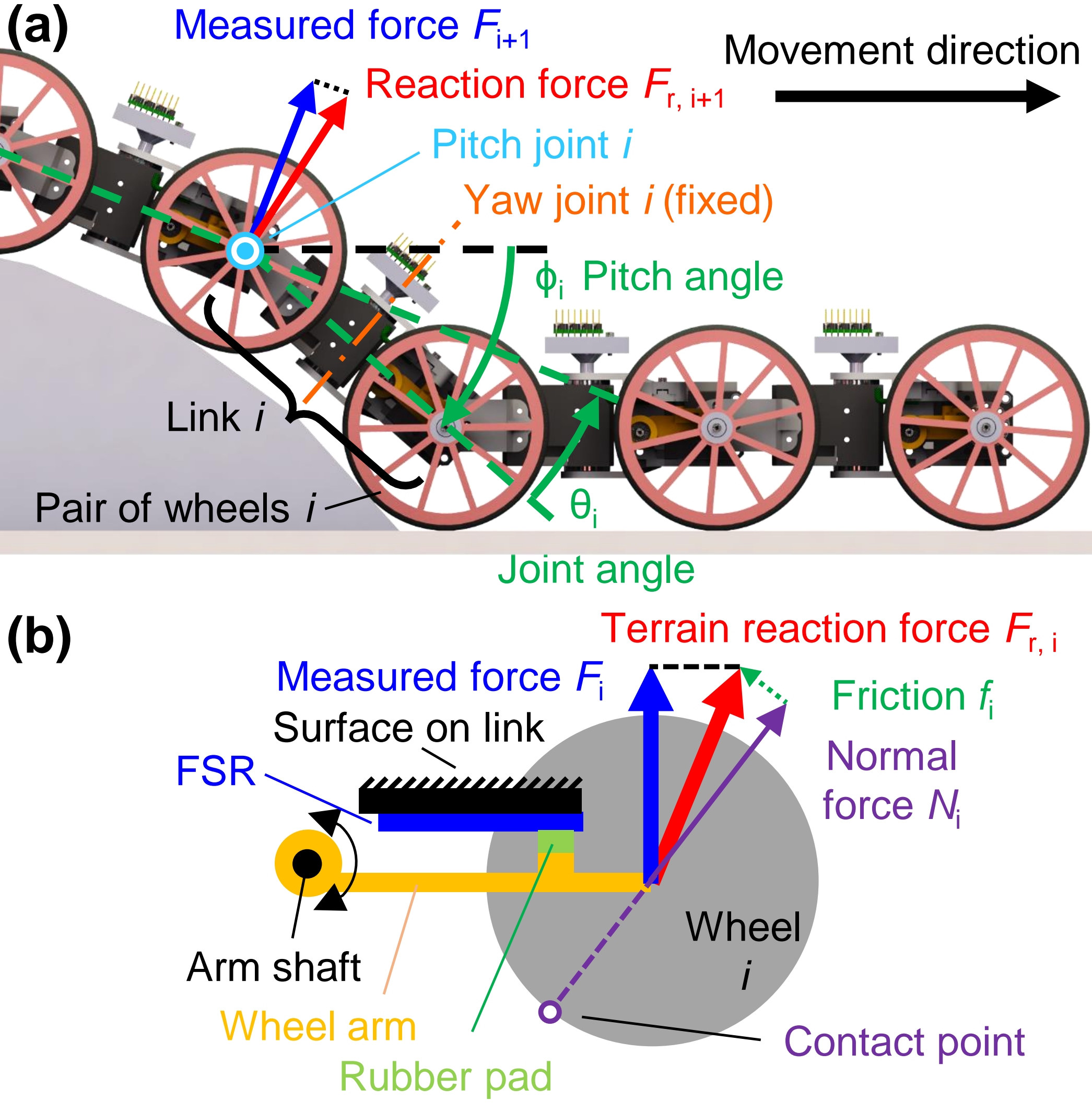

**Figure 2. Design of robophysical model. (a)** Overall structure. Blue and red vectors show measured and total reaction force, respectively, on pair of wheels $i$ + 1. $\phi_i$ and $\phi_{IMU}$ are the pitch angles (positive if clockwise) of link $i$ and the IMU, respectively. **(b)** Installation of force sensing resistor (FSR).

To reduce the number of contact points for easier contact force sensing, a pair of passive wheels (diameter = 87 mm; figure 2(a), red) with ball bearings were installed on the left and right sides at each end of each link. The passive wheels also reduced the fore-aft friction ($\mu = 0.14$, supplementary material section 1.3), which allowed us to test a larger range of backward load as the robot can overcome higher load with less frictional drag. Because of the symmetry about the vertical plane in which the body bent, hereafter we refer to one pair of passive wheels as one wheel for simplicity unless otherwise specified.

To sense terrain reaction forces, we installed a force sensing resistor (FSR; FSR-400 short, Interlink Electronics, Camarillo, CA, USA; figure 2(b), blue) between each passive wheel and the corresponding robot link (supplementary materials sections 1.4). We calibrated the FSRs (supplementary materials section 1.5) and characterized their creep after a sustained constant load or the fatigue after a sequence of loading (supplementary materials section 1.6). While hysteresis and fatigue changed the reading, which is unavoidable for these low cost, small force sensors, we found that the changes were negligible compared to the range of force measured by the sensors during locomotion experiments. To sense the direction of gravity, which is important for controlling a vertically bent body section to push against the terrain in the right direction, we installed an inertial measurement unit (IMU; BNO055 breakout, Adafruit, New York, NY, USA) to the last motor and estimated the pitch angle $\phi_i$ of each link using its reading (supplementary material section 1.2). The readings of the FSRs and the IMU were first collected by a microcontroller board (Mega 2560, Arduino, Turin, Italy) and then sent to the computer at a sampling frequency of 23 Hz.

### 2.2. Controller design

To test the effects of different feedback usage, we implemented five types of controllers to propagate a vertical bending shape posteriorly (figure 3). We first give an overview of these controllers and then explain the biological inspiration of each before introducing the implementations.

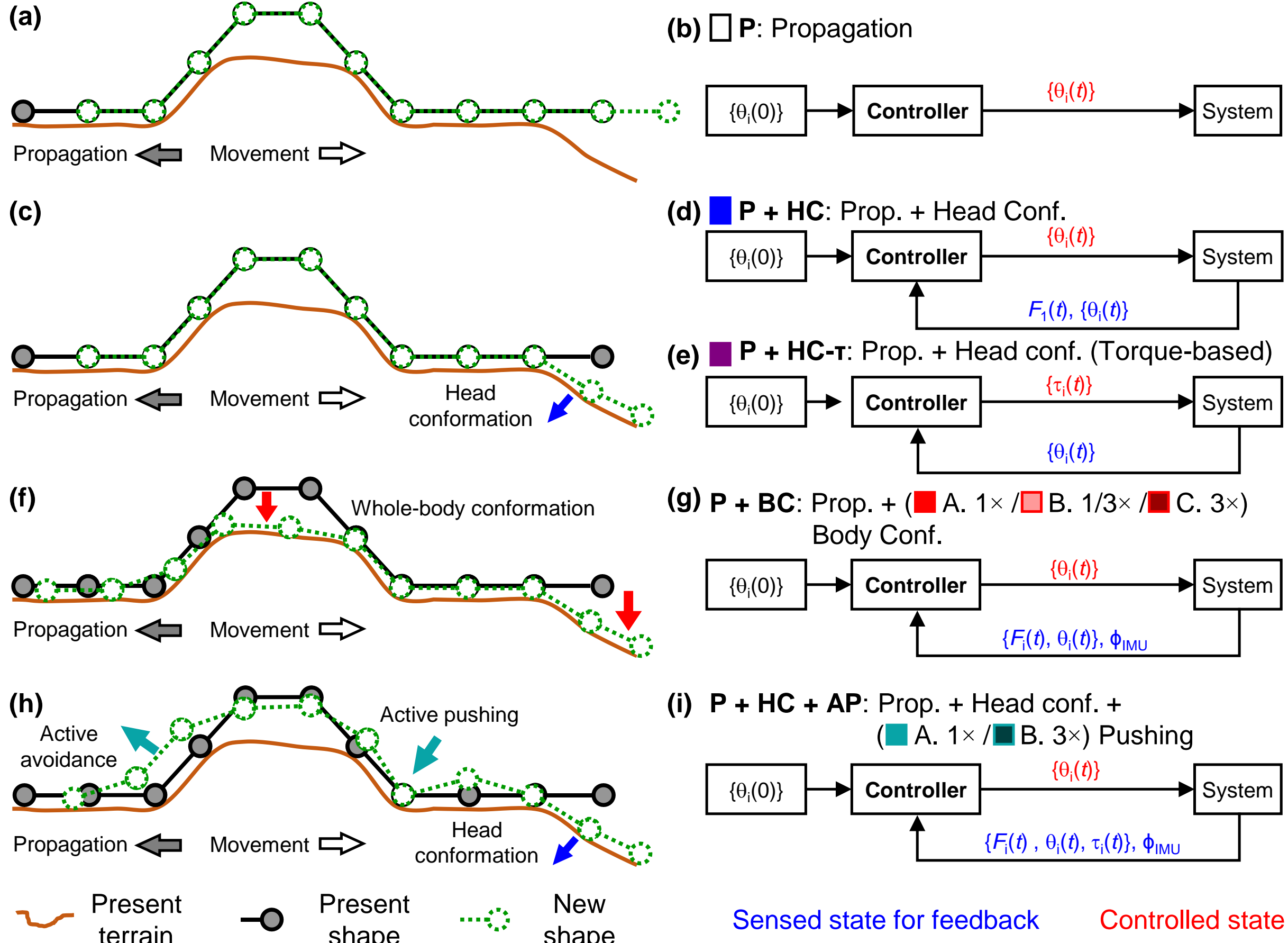

**Figure 3. Comparison of controllers. (a-b)** Controller P: Feedforward shape propagation. **(c-e)** Controllers P + HC and P + HC-τ: Propagation with head conformation using position-based (P + HC) or torque-based (P + HC-τ) control. **(f-g)** Controller P + BC: Propagation with whole-body conformation. **(h-i)** Controller P + HC + AP: Propagation with head conformation and active pushing. Schematics on the left qualitatively show the present robot shape (black solid), the terrain (brown solid), and the expected body bending after using each controller (green dashed). Diagrams on the right show the sensed (blue texts) and controlled states (red texts) used by each controller. A-C in (g) and A, B in (i) represent controllers with different degrees of controlled conformation or active pushing. $F$, θ, τ, and Φ represent measured force, joint angle, joint torque, and pitch angle, respectively.

(1) Controller P is a feedforward controller propagating a pre-determined initial shape posteriorly (figure 3(a, b)).

(2, 3) Controllers P + HC and P + HC-τ control the first link (head) to conform to the terrain and propagate the changes in the body shape posteriorly by sensing the present body shape (figure 3(c)). Controller P + HC controls the head conformation to maintain the sensed contact force at the head and realizes the propagation by controlling the joint angles (figure 3(d)). Controller P + HC-τ realizes both the head conformation and the propagation by controlling the internal force (pitch joint torque) [7] without directly sensing contact (figure 3(e)).

(4) Controller P + BC also uses contact feedback to modulate shape propagation (figure 3(g)). Unlike Controller P + HC that modulates the head bending only, it senses the contact along the body and the gravity direction to control the entire body to conform to the terrain below (figure 3(f)).

(5) Controller P + HC + AP is modified from Controller P + HC to additionally control the body posterior to the first pitch joint to push harder against potentially propulsive contact points and less against potentially resistive ones (figure 3(h)). Specifically, the body section near a push point bends more concavely (figure 3(h), right green vector) if the point is behind this section and bends more convexly (left green vector) if the point is in the front [12]. This controller uses readings of contact forces, joint torques, and joint angles along the body (figure 3(i)) [12].

To further reveal the effects of the additional conformation and pushing during vertical shape propagation, we varied the degree of controlled conformation in Controller P + BC (figure 3(g), A-C) and the degree of active pushing in Controller P + HC + AP (figure 3(i), A, B). Hereafter we refer to Controllers P, P + HC, P + HC-τ, P + BC (A), and P + HC + AP (A) as the five basic controllers.

Besides informing snake robot control, analyses of the performance of the robot using these five types of controllers will inform animal behaviors or sensory mechanisms that inspired these controllers. As an extreme case that does not react to any environmental changes, Controller P served as a control in the comparison. Controllers P + HC and P + HC-τ mimic the follow-the-leader behaviors of a snake traversing uneven terrain accompanied by potential exploration behaviors of the head [15], [31]. Controllers P + BC and P + HC + AP generate the whole-body conformation to or pushing against the terrain, similar to generalist snakes changing lateral bending shape of the entire body to maintain contact with push points [5], [6], [41]. These controllers used multiple sensory information that can be sensed by the snakes, such as contact, body bending, internal force, and gravity [36]–[38].

All the controllers were implemented on the desktop computer and ran at 100 Hz, higher than that of measuring contact force (23 Hz), pitch angle of the IMU (23 Hz), and joint angles and current (31 Hz). To allow sufficient time for the controller to receive and respond to the measurement and protect the robot, the propagation speed and gains of the sensory feedback were set conservatively.

**2.2.1. Feedforward shape propagation**

To propagate the present body shape posteriorly (figure 3(a-b), figure 4), this controller controls the angle of each pitch joint, $\theta_i$, to approach that of the pitch joint in front of it, $\theta_{i-1}$ [44]:

$$\theta_i(t) = \theta_{i-1}(nT) \cdot \frac{t-nT}{T} + \theta_i(nT) \cdot \frac{(n+1)T-t}{T}, nT \leq t < (n+1)T, i \geq 2 \quad (1)$$

where $t$ is the present time, $T$ = 8 s is the time taken to propagate a shape down one link (hereafter referred to as the period), $n$ is the number of periods that have passed, and $i$ is the index of the joint counting from the head.

The first pitch joint angle is controlled to linearly decay to and stay at zero (the first two links being straight) after one period:

$$\theta_1(t) = \begin{cases} \theta_1(0) \cdot \frac{T-t}{T}, t < T \\ 0, t \geq T \end{cases} \quad (2)$$

When using Equations (1-2) together, the initial shape is only propagated down the body once, and then the robot becomes completely straight except for the active wheel module (figure 4). Because the entire bending pattern solely depends on the pre-determined initial shape, Controller P is feedforward despite having the terms $\theta_{i-1}(nT)$ and $\theta_i(nT)$ in Equation (1).

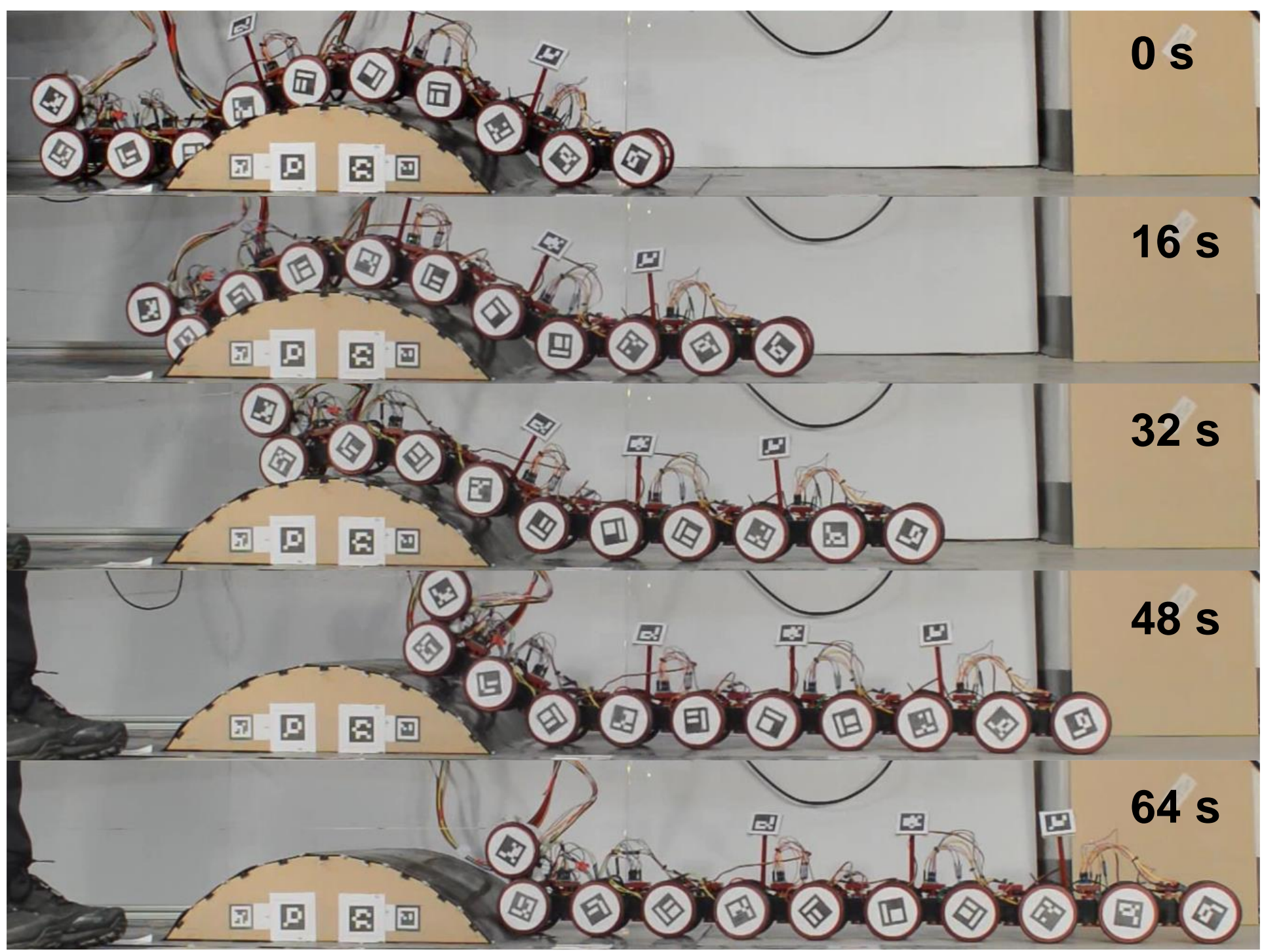

**Figure 4. Robot propagating body shape posteriorly using Controller P.**

**2.2.2. Propagation with head conformation**

To propagate a shape, Controller P + HC also controls the pitch joints behind the head using Equation (1). To control the head to adapt to the terrain when using vertical body bending, previous studies used manual input [28], [29]. Controller P + HC controls the head by maintaining the measured force on the first wheel using a proportional controller with constrained bending velocity:

$$\dot{\theta}_1(t) = \begin{cases} k_1(F_1 - \bar{F}), |F_1 - \bar{F}| < \Delta F_{max} \\ k_1 \Delta F_{max} \cdot sgn(F_1 - \bar{F}), |F_1 - \bar{F}| \geq \Delta F_{max} \end{cases} \tag{3}$$

where $k_1$ = 0.4 rad · s$^{-1}$ · N$^{-1}$ is a constant gain that controls the speed of conformation, $F_1$ is the force detected by the most anterior FSR, $\bar{F}$ = 0.25 N is the desired contact force, and $\Delta F_{\max}$ = 0.25 N is the threshold to limit the speed of conformation. We use a small $\bar{F}$ so that the head does not push against the terrain too hard to avoid large backward resistance from large normal force when pushing against a steep uphill.

**2.2.3. Torque-based propagation with head conformation**

Controller P + HC-τ also controls the head to conform to the terrain while bending each of the other pitch joints toward the angle of its anterior pitch joint. Different from Controller P + HC, the controlled states are the joint torques instead of joint angles. This allows the body shape of the robot to adapt to complex terrain because of additional deformation caused by external forces, such as contact force or gravity, without sensing them [7], [45]. Controller P + HC-τ was adapted from a previous controller [7] which was obtained using a theoretical model traversing a continuous 3-D terrain. The authors optimized a cost function that increases with joint torques assuming that there is no sideslip and no external forces such as friction and gravity [7]. In this study, we controlled the torque $\tau_i$ of the *i*-th pitch joint by controlling the current $I_i$ of the corresponding motor, assuming a linear correlation between them: $\tau_i = k_\tau I_i$, where the torque constant $k_\tau = 1.78$ N · m · A$^{-1}$ [43]. Torque $\tau_i$ was defined to be positive if it was counterclockwise. The previous controller using feedback signal of both joint angles and velocity of the head [7] was also simplified to use only feedback signal of the joint angles:

$$I_i = \begin{cases} I_{head}, i = 1 \\ K_P(\theta_i - \theta_{i-1}), i \geq 2 \end{cases} \tag{4}$$

where $I_{head} = 0.016$ A is a constant current to bend the head toward the terrain and $K_P = -0.53$ A · rad$^{-1}$ is a constant gain.

**2.2.4. Propagation with whole-body conformation**

To allow simultaneous shape propagation and conformation to the terrain, we used a backbone method previously used on a laterally bending robot [42] to guide the bending of the discrete body of the robot more intuitively (figure 5). The backbone method fits the robot's discrete links (figure 5(a), blue) to a continuous virtual curve (backbone curve; figure 5(a), red curve), which goes through a series of virtual shape control points (SCPs; figure 5(a), circles with red outlines). The SCPs are manipulated to deform the curve. The SCPs are initially placed at the endpoints (figure 5(a), red solid circles) of all the robot links (figure 5(a), blue solid). The relative positions of the SCPs are calculated using the joint angle readings and forward kinematics. A cubic spline [46] is then fitted to all the SCPs and used as the backbone curve (figure 5(a), red solid curve).

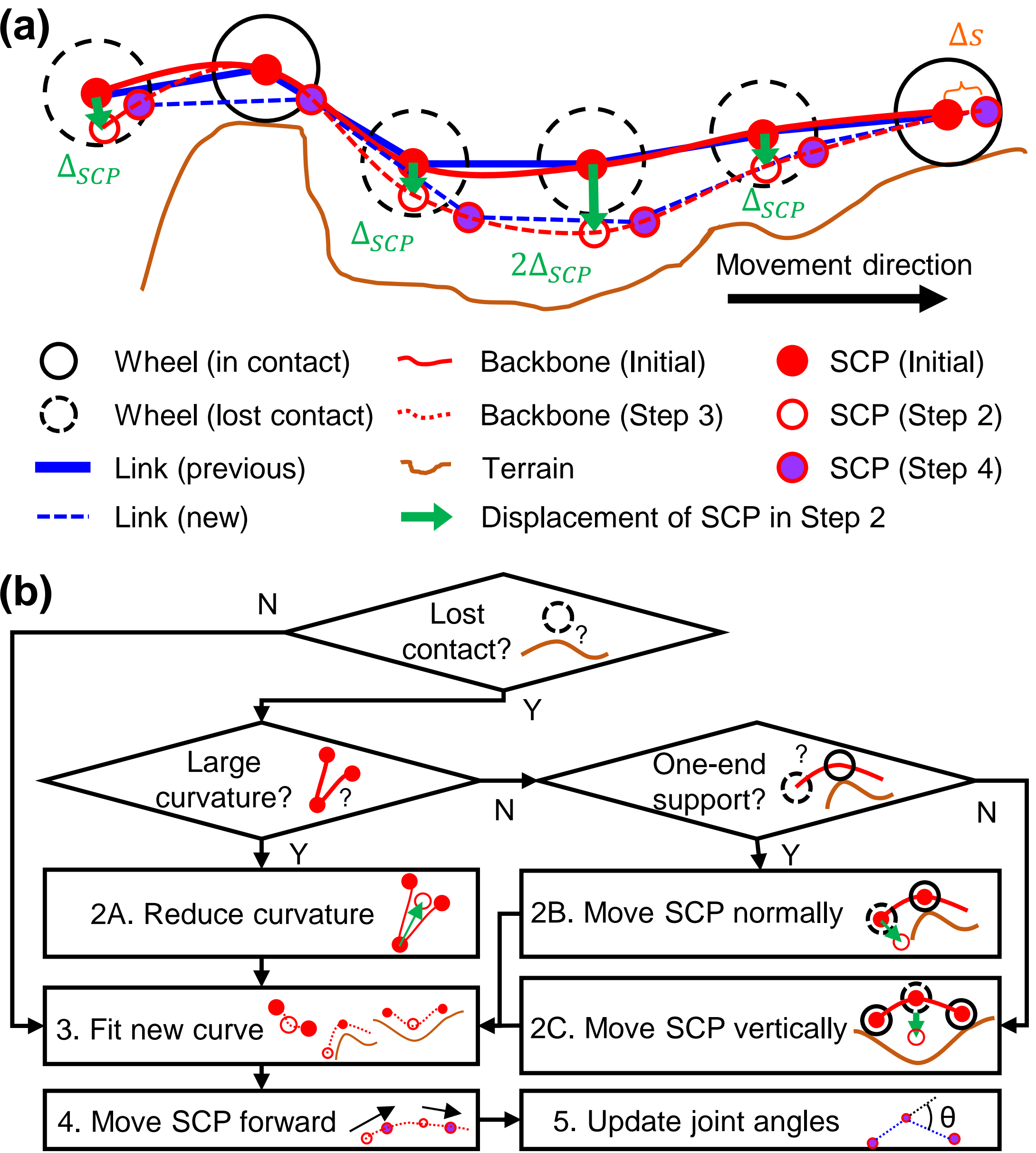

**Figure 5. Implementation of Controller P + BC. (a)** Definition of backbone curve and SCPs. This controller controls body bending by fitting robot links (blue) to a virtual spline curve (backbone curve, red) that is controlled by virtual shape control points (SCPs, circles with red outlines). **(b)** Flowchart of a control cycle.

In each control cycle, the controller updates the joint angles as follows (figure 5(b)): (1) Identify the wheels (figure 5(a), black dashed) that lost contact with the terrain (brown) by checking whether the measured force $F_i$ equals to 0. (2) Move the SCP corresponding to each wheel that lost contact toward the terrain: (A) If the corresponding pitch joint angle already exceeds 60°, move the SCP normal to the local backbone curve toward the concave side by $3\Delta_{SCP}$. This is to avoid collision between different links. (B) Otherwise, if the SCP belongs to a suspended section of the robot with only one end contacting the terrain (figure 5(a), left most red solid circle), move the SCP downward by $n_c \times \Delta_{SCP}$ normal to the line segment connecting it and the nearest SCP whose wheel is contacting the terrain (figure 5(a), left most green arrow), where $n_c$ is the number of links between this and that nearest SCP. (C) Otherwise, move the SCP vertically downward by $n_c \times \Delta_{SCP}$ (figure 5(a), the right three green arrows). (3) Fit a new cubic spline curve (figure 5(a), red dashed) to the updated SCPs as the new backbone curve. (4) To propagate the shape posteriorly, move the most anterior SCP forward along the new backbone curve by $\Delta s$ (figure 5(a), right most red circle with purple filling), then move the other SCPs along the new backbone curve such that each line segment connecting two adjacent SCPs (figure 5(a), blue dashed) has the same length as a robot link. (5) Calculate the new joint angles between the adjacent line segments fitted in (4). In this study, we used $\Delta s = 0.2$ mm in a control cycle period of 0.01 s. We used $\Delta_{SCP} = 0.1$, 0.035, or 0.3 mm to apply a medium, small, or large degree of controlled conformation in the three variants A-C, respectively. In this controller, force sensors are used as on-off switches that only detect whether each wheel contacts the terrain or not.

**2.2.5. Propagation with head conformation and active pushing**

Controller P + HC + AP was adapted from a controller previously proposed for lateral bending [12], [28]. Aside from reducing the pushing against obstacles in front of the body similar to the controllers using controlled compliance [2], it can help the robot actively push harder against push points behind the body. Different from the original design, Controller P + HC + AP propagates body bending in the vertical plane using position control instead of torque control and further considers gravity (figure 3(h, i)):

$$\dot{\theta}_i = \dot{\theta}_{i,w} + \dot{\theta}_{i,p}, \dot{\theta}_{i,p} = \sigma_i \cdot tanh\left(\alpha \sum_{j=i-n_a}^{i+n_p} \left(-\tau_j\left(-F_{j+1} + G_j \, cos\left(\phi_{j+1}\right)\right)\right)\right), i \geq 2 \qquad (5)$$

where $\dot{\theta}_{i,w}$ is the change of pitch to propagate a shape calculated using Equation (1), $\dot{\theta}_{i,p}$ is the change of pitch to realize the additional active pushing, $\sigma_i$ controls the speed of this additional change of pitch $\dot{\theta}_{i,p}$, $\alpha = 0.08\ N^{-2} \cdot m^{-1}$ is a gain that controls the sensitivity of $\dot{\theta}_{i,p}$ to force and torque readings, $n_a = 0$ and $n_p = 2$ are the numbers of links anterior and posterior to link $i$, respectively, used for feedback control, $\tau_j$ is the torque output of pitch joint $j$, $F_{j+1}$ is the force detected by the ($j$ + 1)-th FSR, and $G_j$ = 2.94 N for $j$ = 1, …, 9 or $G_j$ = 5.88 N for $j$ = 10 (because of the additional active wheel module) is the weight of one link that we added to account for the effect of gravity in the vertical plane.

To explore how the active pushing affects the performance when used together with vertical shape propagation, we used $\sigma_i$ = 0.013 or 0.039 rad · $s^{-1}$ to apply a small or large degree of active pushing in the two variants A and B, respectively. Note that Controller P + HC is a special case of Controller P + HC + AP when $\sigma_i$ = 0, which applies no additional pushing.

Different from the previous studies that relied on manual control of the head [12], [28], here the first pitch joint was controlled using Equation (3) to conform to the terrain automatically.

**2.3. Experimental design**

In the previous study [15] (figure 1(b)), the snake did not use vertical bending alone until it gained substantial contact with the sloped surface of the wedge, because it needs substantial contact with the vertical push point to use vertical bending to gain sufficient contact force from it for propulsion [47]. Similarly, we provided our robot with the downhill of a bump as an initial vertical push point (figure 6(a), supplementary material section 2.1). We set the initial position such that the initial shape matched the entire terrain geometry, which contained two horizontal sections and a curved section from the bump (figure 6(a)). The robot can traverse this terrain from a more posterior position using Controllers P + HC, P + BC (B), and P + HC + AP (A) but not others (Movie 4). Thus, we used this initial position for all the controllers for direct comparison. The robot reached this position using its active wheel (supplementary material section 2.2). We did not use lateral bending such as concertina (figure 1(b)) or lateral oscillation [13] as snakes do, because these gaits are often accompanied by uncontrollable slipping that prevents repeatable experiments with identical initial conditions.

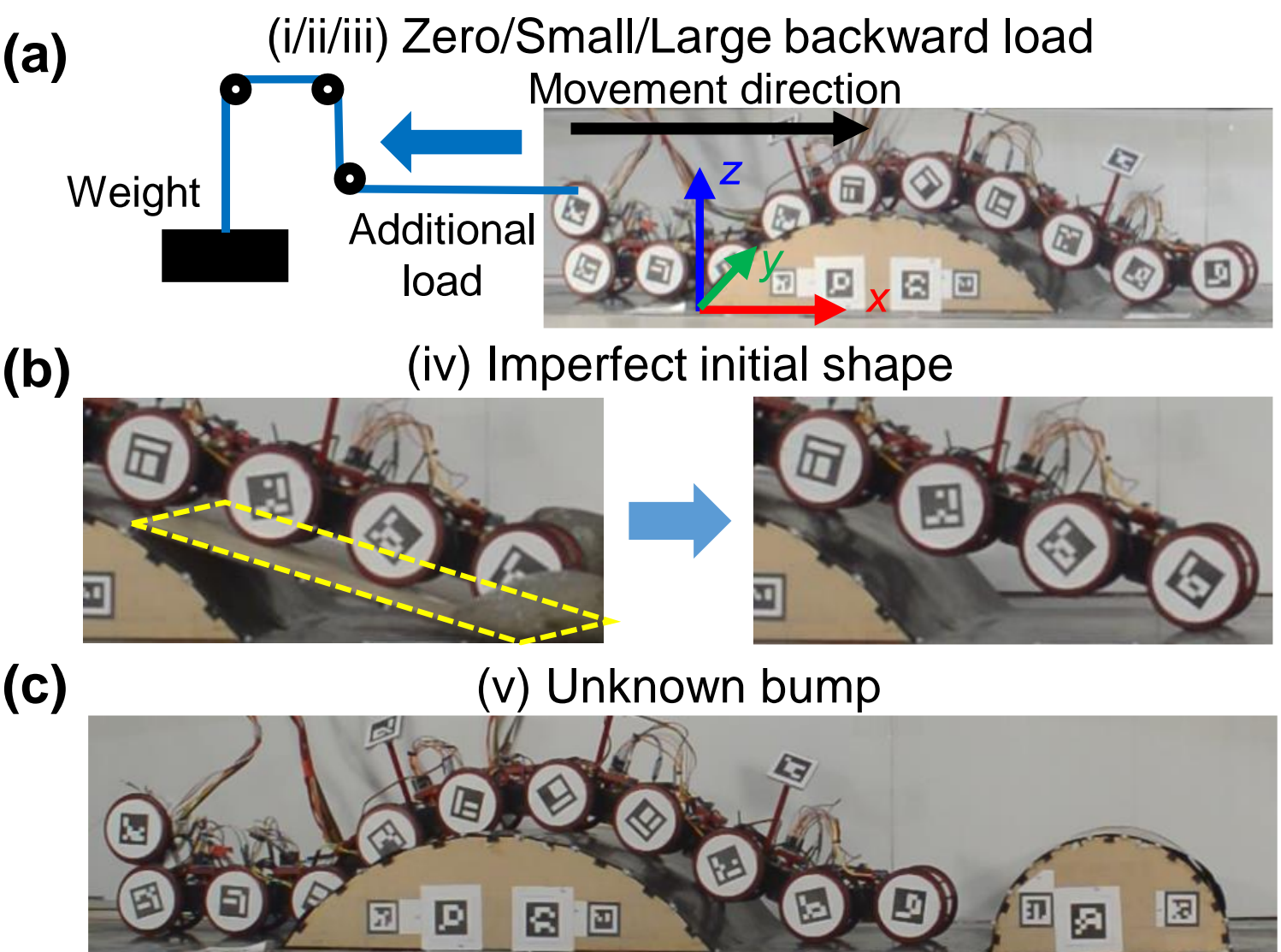

**Figure 6. Experimental setup in different cases. (a)** Traversal with an additional backward load (cases (i-iii)). Load is applied by a weight via a pulley system. **(b)** Traversal with an imperfect initial shape that has poor contact with terrain (case (iv)). An acrylic plate (yellow dashed) is placed under robot before initial body shape is determined and removed before a trial starts. **(c)** Traversal with an unknown bump in front (case (v)).

To apply different backward loads in cases (i-iii), we connected the tail of the robot to slotted weights via a string through a pulley system (figure 6(a), supplementary material section 2.3), similar to a drawbar test in terramechanics [48]. The three loads in cases (i-iii) were 0, 1.5, and 2.9 N (0%, 36%, and 71% of the backward friction on the robot lying straight on flat ground), respectively.

To study whether the robot can move forward using vertical bending without contacting a push point initially and how well it can regain this lost contact in case (iv), we added an acrylic plate below (figure 6(b), left), let the robot conform to the plate and recorded initial joint angles, and removed the plate (figure 6(b), right) before a trial began (supplementary material section 2.3).

To test how well the robot can conform to novel obstacles in front of it in case (v), we added an additional 0.13 m high, 0.25 m long half-cylindrical bump 0.29 m in front of the main bump (figure 6(c), supplementary material section 2.3).

**2.4. Data collection and analyses**

To compare the performance of different controllers, we challenged the robot to traverse the track using each of the 8 controllers for each of the 5 cases with a variation of backward load or terrain geometry, each with 5 trials, resulting in a total of 200 trials. We recorded the kinematics of the robot by tracking the ArUco markers [49] attached to each wheel, which were captured by four synchronized cameras at a sampling frequency of 60 Hz [50]–[52] (supplementary materials section 2.4). We also reconstructed the 3-D terrain profile for evaluating contact conditions by tracking the ArUco markers attached to the terrain (supplementary materials section 3.1.2). We recorded FSR forces, present motor angles and current, motor goal angles and current that were computed by each controller, and IMU orientation, and synchronized them with the kinematics obtained from the cameras (supplementary materials section 2.4).

We quantified the performance of the robot when using each controller in each case in 4 different aspects (figure 7; supplementary material section 3.1): (1) overall performance, measured by success rate, defined as the ratio between the number of successful trials and that of all the trials for each controller in each case, (2) actual terrain conformation, measured by the spatiotemporal average of clearance, which was defined as the closest distance between each passive wheel and the terrain [53], (3) demand on the actuators, measured by the maximal torque generated by a pitch joint motor (excluding the one lifting the active wheel) during a trial, and (4) demand on the power supply, measured by the maximal total current consumed by all the pitch joint motors (excluding the one lifting the active wheel) in a trial.

To understand how modulation of body bending from feedback control affected body-terrain contact, we calculated the contact forces (figure 2(b), red), including the normal contact force (purple) and the friction (green) on all the passive wheels, using the tracked kinematics and the measured FSR forces (supplementary material section 3.2). Then, we analyzed the dependence of contact force on the number of suspended links and the pitch joint torque. To understand how body-terrain contact affected performance, we also analyzed the effect of the slope of terrain surfaces that the robot contacted on propulsion generation. See details of these analyses in supplementary material section 3.3.

## 3. Results

Testing various backward loads and terrain geometries revealed differences in how different controllers helped the robot bend differently under unexpected challenges. These differences further led to different body-terrain interaction which affected the performance. Analyzing the results helped us test our hypotheses: (1) When only propagating a vertical bending shape posteriorly, the robot can generate propulsion to traverse the track under zero backward load (i) and accommodate a small or large backward load (ii-iii). (2) When using feedforward control, the robot will struggle more under terrain variations (iv-v) than when without such variations (i-iii). (3) When using contact feedback control, the robot can better accommodate terrain variations than when using feedforward control.

Below, we first describe the performance of the five basic controllers in each case (Section 3.1) and discuss failure modes (Section 3.2). Then we analyze how bending patterns affected system performance and how sensory feedback control modulated them by modulating contact conditions (Sections 3.3-3.5).

### 3.1. Performance of the five basic controllers

#### 3.1.1. Success rates and bending patterns

Using Controller P, the robot achieved high success rates and maintained good contact with the terrain when the backward load was zero or small (figure 7(a-b) and Movie 1, i-ii). However, it failed in 20% of the trials when the backward load was large and in all the trials when terrain geometry varied (figure 7(a), iii-v). In the failed trials, the robot always lost substantial contact with the terrain, which resulted in large clearance (figure 7(b) and Movie 1, iii-v).

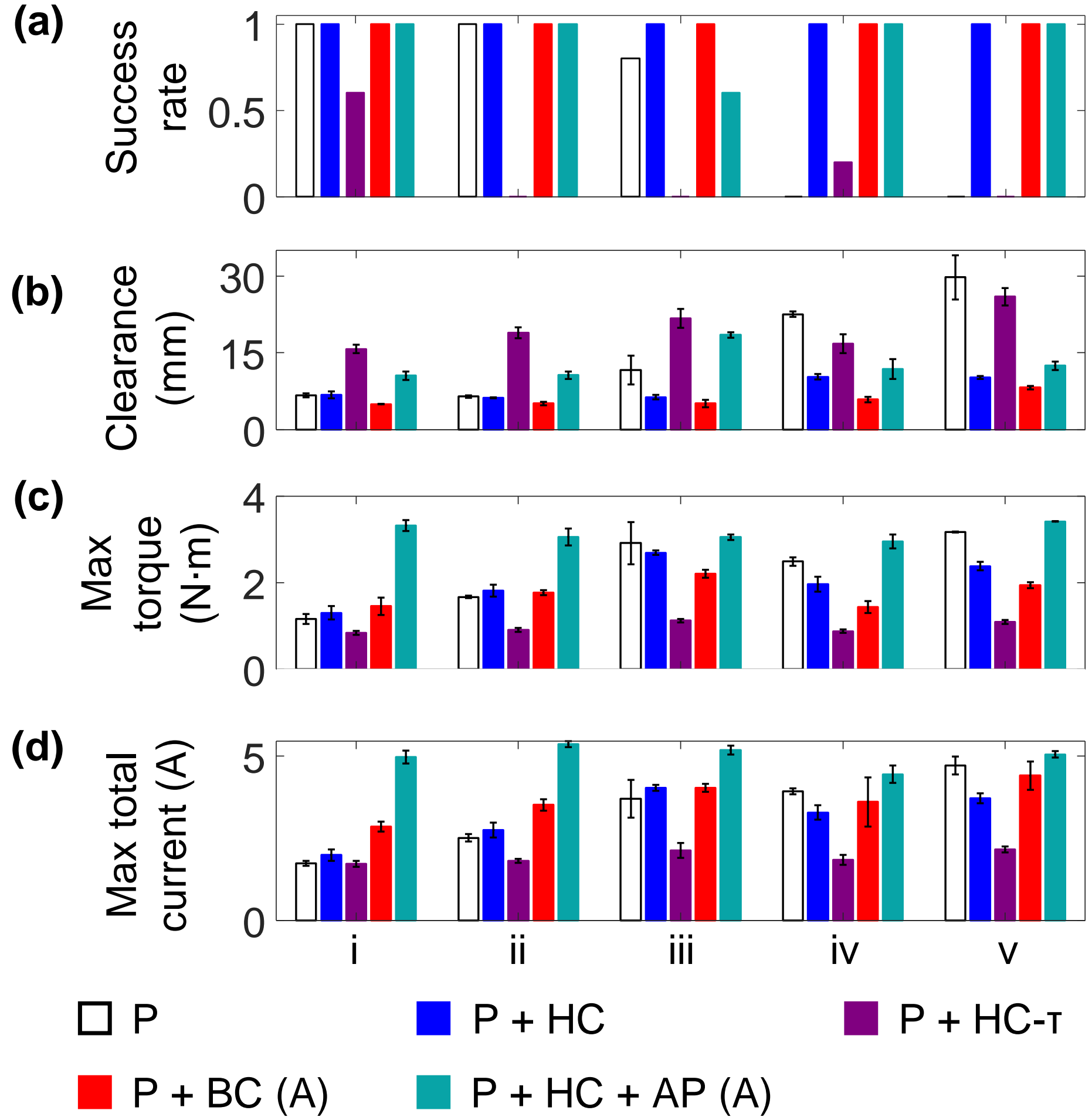

**Figure 7. Comparison of performance. (a)** Overall performance, quantified by success rate. **(b)** Actual terrain conformation, quantified by average clearance of all passive wheels. **(c)** Demand on actuators, quantified by maximal torque generated by pitch joints. **(d)** Demand on power supply, quantified by maximal total current consumed by pitch joints. Error bars show ± 1 s.d.

Using Controllers P + HC and P + BC (A), the robot succeeded in all the trials (figure 7(a)) and maintained good contact with the terrain regardless of the backward load or terrain variation (figure 7(b), Movie 1).

Using Controller P + HC-τ, the robot failed in 40% of the trials when the backward load was zero, in 80% of the trials when starting with an imperfect initial shape, and in all the trials when the backward load was small or large or when there was an unknown bump (figure 7(a)). When the load was large, the robot slipped backward immediately after the controller started (Movie 1, iii). In the failed trials in the other cases, it initially moved forward rapidly and kept accelerating, but it stopped moving forward after the last passive wheel contacted the bump (Movie 1, i-ii and iv-v). Controller P + HC-τ always caused large up and down oscillations at each pitch joint (Movie 1), which resulted in a large average clearance (figure 7(b)).

Using Controller P + HC + AP (A), the robot succeeded in all the trials regardless of backward loads or terrain variations, except in two trials when the backward load was large after motor stalling (figure 7(a)). The robot always lifted part of its body section in front of the main bump (Movie 1), resulting in large clearances (figure 7(b)).

**3.1.2. Maximal joint torque and maximal total current**

When the robot used Controllers P, P + HC, and P + BC (A), increasing the backward loads increased maximal pitch joint torque and maximal total current consumed by pitch joints (figure 7(c, d), i-iii). The variations of terrain geometry also increased the maximal torque and maximal total current compared to when there was no backward load (figure 7(c, d), iv-v versus i), except for the maximal torque when the robot used Controller P + BC (A).

When using Controller P + HC-τ, the robot consistently generated smaller maximal torque and used smaller maximal total current than when using the other controllers regardless of the backward load or variation of terrain geometry (figure 7(c-d)).

When using Controller P + HC + AP (A), the robot consistently generated large pitch joint torque exceeding the capacity of the servo motors (figure 7(c)) and consumed large total current (figure 7(d)) in all the cases. This stalled the fifth or the sixth pitch joint motor in all the trials except two trials in case (iv) with an imperfect initial shape (Movie 1). Regardless, the robot managed to complete the traversal in most of the trials (figure 7(a)).

**3.2. Failure modes**

There were three failure modes among the failed trials: (F1) stop of forward movement despite accurate shape propagation (figure 8(a)), (F2) stop of shape propagation because of motor stalling (figure 8(b)), and (F3) stop of forward movement after distortion of the body shape being propagated (figure 8(c)).

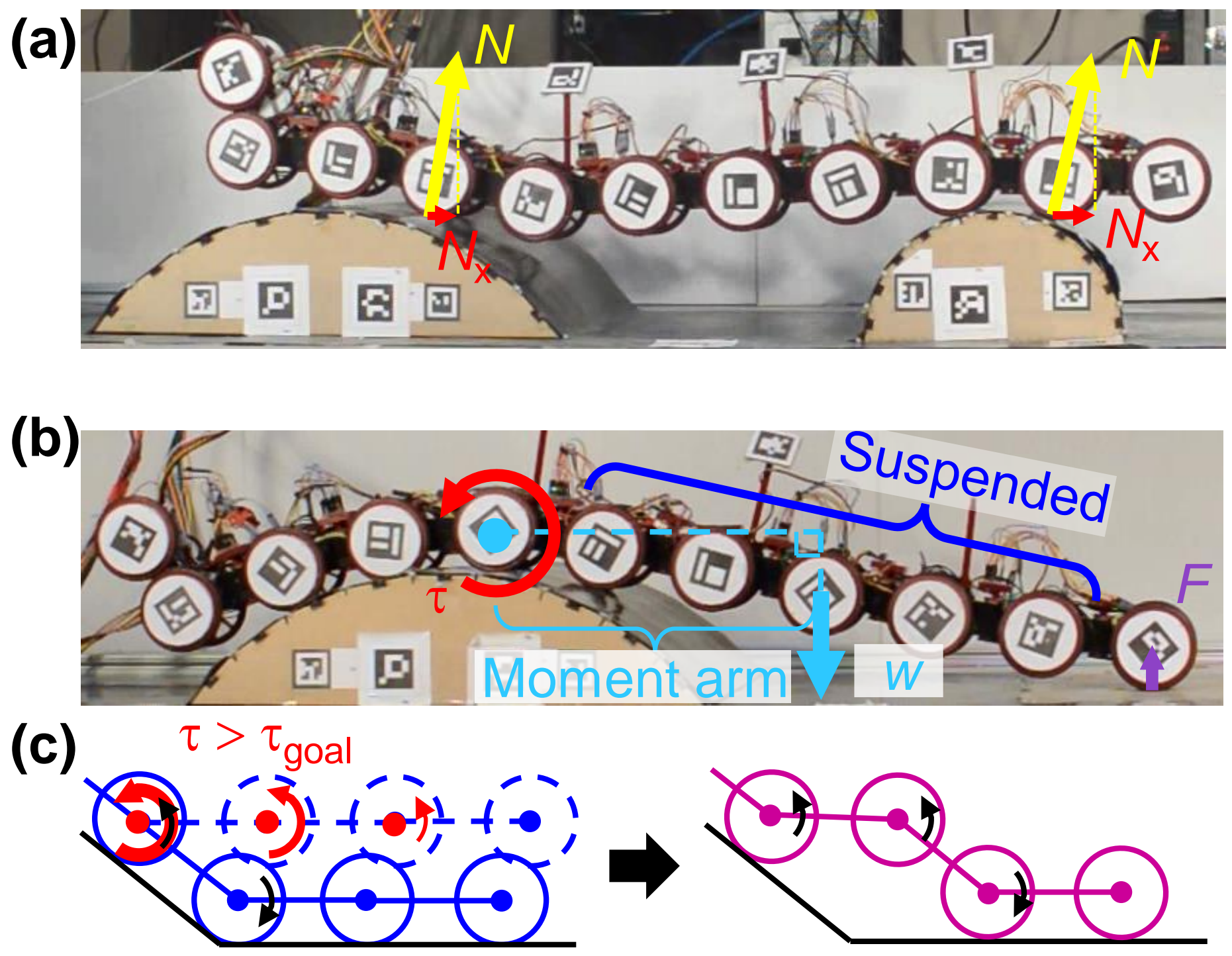

**Figure 8. Typical failure modes. (a)** F1: Stop of forward movement despite accurate shape propagation. Limited propulsion can be generated from horizontal components (red) of normal contact forces (yellow). **(b)** F2: Stop of shape propagation because of motor stalling. A large counterclockwise pitch joint torque τ (red) is required to balance the clockwise torque about the same pitch joint (cyan dot) from the weight *w* of the suspending body section (cyan arrow), especially when limited counterclockwise torque is provided by the terrain reaction force *F* of the other end (purple). **(c)** F3: Stop of forward movement after distortion of body shape. Blue solid lines and circles show present shape of robot, blue dashed lines and circles show desired shape after propagation, and magenta shows actual shape after propagation. Red and black arrows show expected torque τ to apply large contact forces on the downhill by propagating shape and goal torque $\tau_{goal}$ generated by controller, respectively.

When using Controller P, the robot failed from modes F1 and F2 when the backward load was large or there were terrain variations. After losing contact with the steep downhills (figure 8(a)), the forward components (red) of the normal contact forces (yellow) were insufficient for overcoming resistance from friction and pushing against the uphills. When a substantial part of the body was suspended in the air (figure 8(b), blue bracket), overcoming the clockwise torque from the large weight (cyan) required a large counterclockwise torque at the pitch joint lifting this section (red), which often stalled the motor. When using Controller P + HC + AP (A), the robot failed from mode F2 in 40% of the trials in case (iii) with a large backward load because of the same reason.

When using Controller P + HC-τ, the robot failed from mode F3 in all the cases. To drag the heavy tail up the bump, to overcome the backward loads in cases (ii) and (iii), or to propel up the additional bump in case (v), the robot must generate a large forward component of contact force by pushing the wheels normally downward against the downhill. This increased the vertical contact force on these wheels (supplementary material section 2.4) and reduced that on the other wheels in front of the bump, because the sum of vertical contact forces on all wheels must balance the weight. This often lifted the anterior wheels off the terrain (Movie 1, iii; figure 8(c), from blue solid to blue dashed), which required that the pitch joints in front of the bump (figure 8(c), red) generate large torques to maintain the shape. Such large torques often exceeded the goal torque computed by Equation (4), which was zero if the suspended body section was straight. Thus, the body shape was deformed downward by the weight of the suspended sections (figure 8(c), right), the propagation was stalled, and the robot could not push the push points sufficiently hard to move forward. This was also reflected by the smaller maximal pitch joint torque than that when the robot used the other position-based controllers (figure 7(c)).

We tried increasing the gain $K_P$ in Equation (4) to increase the goal torque (Movie 2). This induced larger and faster up and down oscillations that damaged the robot in most of the attempted trials, despite an increased success rate in the other trials. We could not eliminate these oscillations by introducing integral and derivative terms in the feedback controller. We suspect that the persistent oscillation resulted from three issues: (1) The relationship between the output torque and the motor current has hysteresis for our motors with a large gear ratio [54]. (2) The actual feedback control frequency was slow due to the slow sampling frequency of the joint angles (31 Hz), which could result in a lag between the sampling and the controlled output. This issue also likely led to the small up and down oscillations of the pitch joints when the robot used Controllers P + HC and P + BC (A) (Movie 1). The effect of the lag was larger on Controller P + HC-τ, using which the robot kept bending each joint with a constant torque in each control loop, than the other controllers that only bent each joint with an increment designed to be small (Equations 1-3 and 5). (3) The development of the controller did not consider gravity, longitudinal friction, and additional backward load [7]. Thus, the torque generated by one joint (Equation 4) only depends on the present body shape. This resulted in not only the insufficient pitch joint torque to hold a straight suspended body section (figure 8(c)), but also the large oscillation and overshoot of pitch joint angles for those suspended in the air when the gain $K_P$ was increased to generate larger torque for pitch joints above the downhills.

### 3.3. Propulsion generation of shape propagation posteriorly in the vertical plane

The high success rate of the robot under no backward load (figure 7(a), i) when using Controllers P, P + HC, P + BC (A), or P + HC + AP (A) supported the hypothesis that vertical bending alone can help the robot traverse the uneven track, and presumably other similar continuous terrain with large height variation which allows a vertically bent body to push against vertical push points. The high success rate of the robot under a small or large backward load when using Controller P + HC or P + BC (A) (figure 7(a), ii-iii) demonstrated that the propulsion generated can also increase to accommodate additional resistance if contact is maintained.

We also observed that vertical body bending enabled the robot to traverse novel terrain if the following conditions on the bending patterns and contact conditions were satisfied: (1) the robot continuously propagates a vertical bending shape down the entire body, (2) the robot maintains contact with steep downhills, and (3) there are sufficient contact points to support the body weight.

The robot succeeded in all the trials when these conditions were met, such as when using Controllers P + HC and P + BC with variations of terrain geometry (figures 7(a, b), iv, v). This was achieved by creating an asymmetry in the body-terrain interaction: as the shape was propagated posteriorly, the body continuously pushed against the downhills for propulsion while detaching the uphills to reduce resistance.

If the robot failed to meet condition 1, namely stopping the shape propagation posteriorly, it stopped moving forward such as in failure mode F3 (figure 8(c)) and after failure mode F2. If the robot failed to meet condition 2, namely losing contact with steep downhills, it failed from failure mode F1 (figure 8(a)).

If the robot failed to meet condition 3, namely a long body section losing contact, it failed from failure mode F2 (figure 8(b)).

### 3.4. Modulation of propulsion by controlling contact conditions using body bending

To further understand why the distinct bending patterns generated by different controllers led to different traversal performance, we analyzed how the bending patterns modulated the body-terrain contact and how the modulation in contact affected traversal performance. We identified three mechanisms in which feedback controllers can modulate the distribution of contact forces and propulsion by changing bending patterns.

#### 3.4.1. Increase of vertical contact force by lifting adjacent body sections

The first mechanism is to push one wheel harder against the terrain below by lifting the body sections adjacent to it off the terrain. The weight of a suspended body section is supported by the contact forces from the terrain and constraint forces from other body sections on its one or two supported ends. In an ideal situation where no constraint forces from other body sections are present and the movement is quasistatic (figure 9(a)), the sum of the vertical contact forces $F_{\mathrm{r,\,z}}$ on the two ends is proportional to the number of suspended links. Although in reality these idealized conditions usually cannot be satisfied, we found from the measured contact data that statistically this linear correlation was significant for the longest suspended body section in each video frame ($P < 0.0001$, $r^2 = 0.407$, 689287 d.f., linear regression; figure 9(b) left). This implies that a robot with contact sensors can deliberately increase the vertical contact force on one wheel by lifting more adjacent body sections off the terrain. For example, when using Controller P + HC + AP (A), the robot lifted many links in front of the bump to push harder against it (Movie 1).

Robots without contact force sensors can potentially identify suspended body sections by checking pitch joint torques and apply this strategy. In the simplified example, the pitch joints within this section need to generate negative (clockwise) torque (figure 9(a), blue) to hold the suspended links. When using this criterion to determine the longest suspended section, the sum of $F_{\mathrm{r,\,z}}$ on both ends was still significantly linearly proportional to the number of suspended links ($P < 0.0001$, $r^2 = 0.289$, 379889 d.f., linear regression; figure 9(b) right).

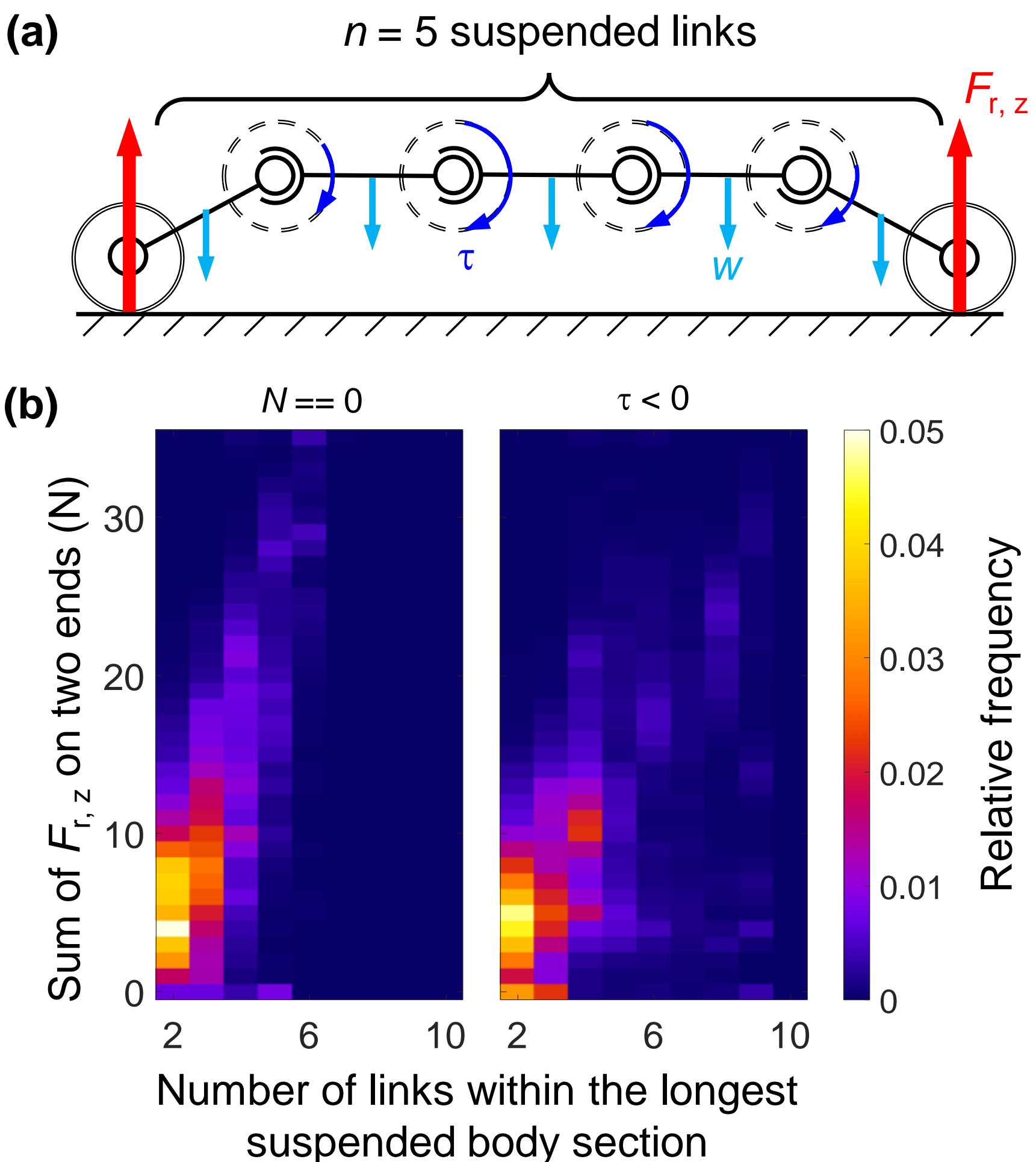

**Figure 9. Relationship between the number of links within a suspended body section and the sum of vertical contact forces on its two ends. (a)** Schematic of a suspended body section in an ideal situation with no constraint forces from other body sections. $F_{r,z}$ shows vertical terrain reaction force, $\tau$ shows pitch joint torque (positive if counterclockwise), $w$ represents weight of a link, and $n$ represents the number of suspended links. **(b)** Normalized histogram of the sum of vertical contact forces exerted by the longest suspended body section, determined by identifying wheels with no contact forces (left) or pitch joints with negative torque (right), and the number of links inside this section. The color of a region shows the relative frequency of the measured values occurring in the corresponding range of values, defined as the number of occurrences divided by the total number of data points.

### 3.4.2. Increase of vertical contact force by increasing local pitch joint torque

Aside from lifting more adjacent links, the robot can increase the vertical contact force on one wheel by increasing the torque of the local pitch joint. Consider an ideal situation where the robot is only contacting the terrain on three wheels, is moving quasi-statically while being horizontal and straight, and is generating positive (counterclockwise) torque on the pitch joint near the middle wheel (figure 10(a)). Increasing positive torque $\tau$ increases the local vertical contact force $F_{r,z,1}$ if the numbers of suspended links adjacent to the wheel do not change (supplementary material section 3.3). Although in reality these idealized conditions usually cannot be satisfied, we found from the measured joint torque and contact force data that, for the pitch joint that was generating the largest torque, the vertical contact force on the wheel closest to this joint statistically increased with the torque of this joint ($P < 0.0001$, $r^2 = 0.224$, 588150 d.f., linear regression; figure 10(b)).

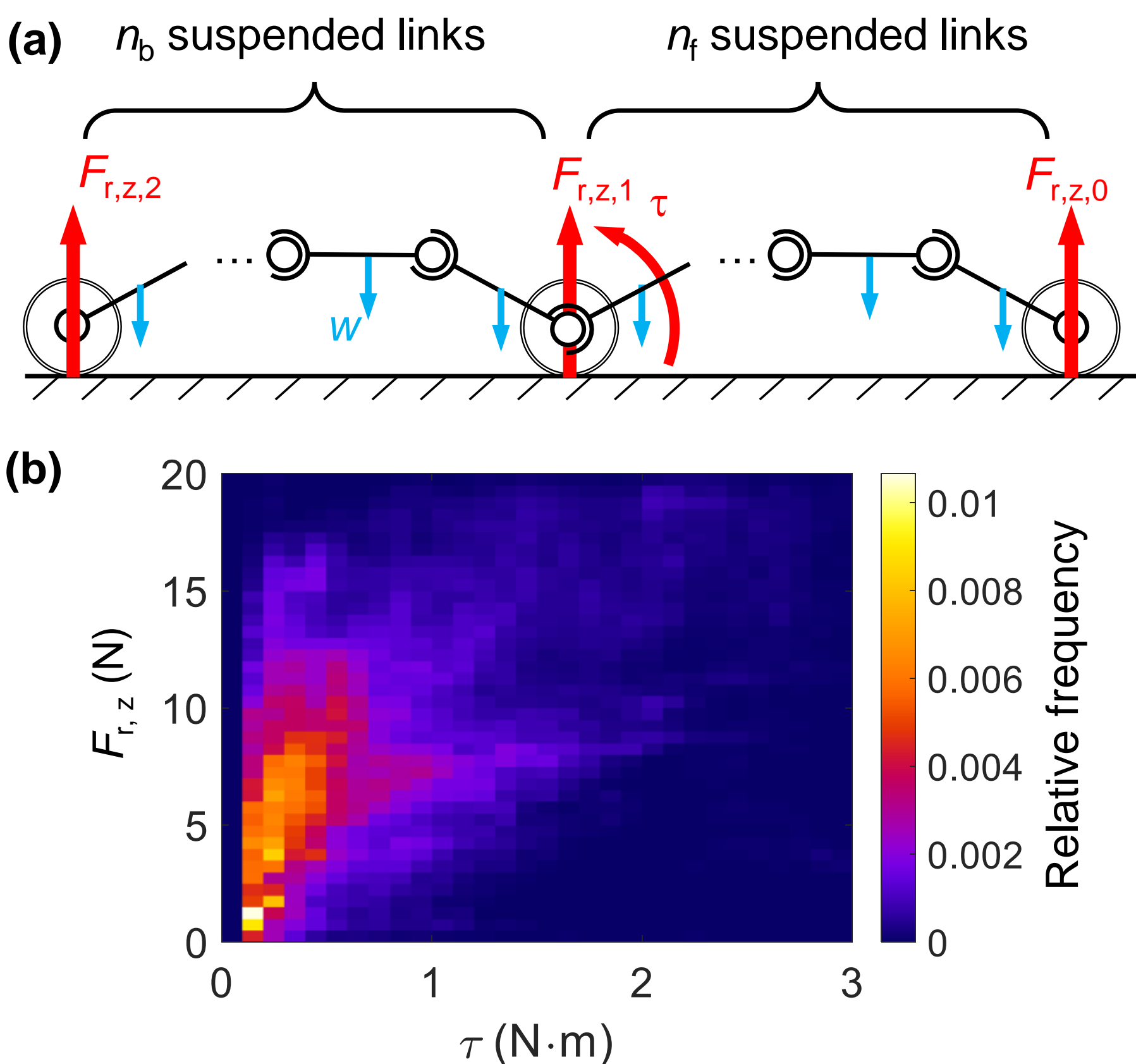

**Figure 10. Relationship between vertical contact force on one wheel and the positive torque generated by the nearest pitch joint.** **(a)** Schematic of the robot contacting the terrain with three wheels

and bending the middle wheel toward the terrain. $F_{r, z, i}$ shows vertical contact force on a wheel ($i$ = 0, 1, 2), τ shows pitch joint torque (positive if counterclockwise), $w$ represents weight of a link, and $n_b$ and $n_f$ represent the number of suspended links on two sides of the middle wheel. **(b)** Normalized histogram of the vertical contact force on a wheel and the bending torque of the nearest pitch joint, for the pitch joint that was generating the largest torque. The color of a region shows the relative frequency of the measured values occurring in the corresponding range of values, defined as the number of occurrences divided by the total number of data points.

### 3.4.3. Effective use of normal contact force to propulsion by pushing against steep downhills

While increasing vertical contact force on a wheel can linearly increase the horizontal contact force (propulsion) on the same wheel if the surface in contact does not change (supplementary material section 3.2), this increase was sometimes insufficient for the robot to move forward. For example, using Controller P, the robot facing an unknown bump developed large vertical contact forces on a few wheels but still cannot move forward (figure 8(a)).

Besides increasing vertical contact forces, which cannot exceed the weight of the entire robot because of the force balance in the vertical direction, increasing the ratio between horizontal and vertical contact forces can help the robot increase propulsion. This can be achieved by pushing against steeper downhills (figure 11(a)). In our locomotion experiments, the downhill from which the robot developed the largest vertical contact force was significantly steeper when the robot was moving than when it was stuck in place (–24.4° versus –15.9°; $P < 0.0001$, $F_{1,715707} = 53793$, ANOVA; figure 11 (b), left). In addition, the steepest downhill that the robot was contacting was also steeper when the robot was moving than when it was stuck (–33.1° versus –31.0°; $P < 0.0001$, $F_{1,693621} = 4052$, ANOVA; figure 11(b), right).

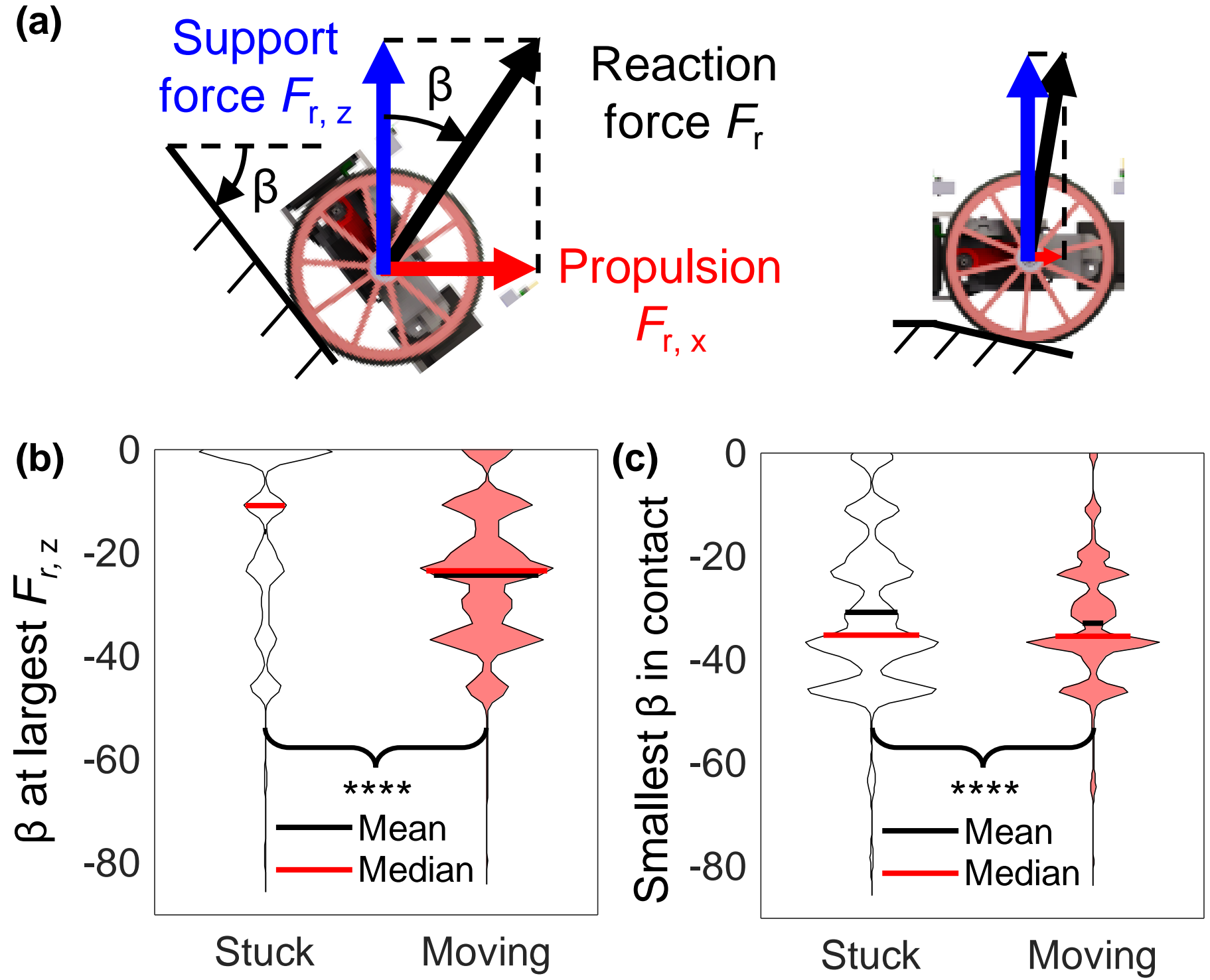

**Figure 11. Benefits of pushing against steep downhills. (a)** Schematic of a wheel pushing against steeper (left) and less steep (right) downhills. Slope of surface β is negative for downhills. **(b-c)** Comparison of slopes when the robot is stuck (white) and when it is moving (red) for: (b) downhill that the wheel with the largest vertical contact force contacts and (c) the steepest downhill the robot is contacting. Data are shown using violin plots. Black and red lines show mean and median, respectively. Local width of graph is proportional to the probability density of data along the y-axis. Brackets and asterisks show a significant difference (****$P$ < 0.0001, ANOVA).

### 3.5. Impact of feedback-controlled conformation and active pushing on performance

Next, we investigated how feedback-controlled conformation to or active pushing against the terrain affected the four performance metrics.

To understand the effect of feedback-controlled conformation, we compared the performance of the robot when using Controllers P and P + BC (A-C), which have a zero, medium, small, or large degree of controlled conformation to the terrain in addition to the propagation, respectively. In the challenging cases (iii-v) with large backward loads or variation of terrain geometry, both Controllers P + BC (A) and

(B) increased the success rate compared to Controller P (figure 12(a)), presumably by improving actual terrain conformation (figure 12(b)) and reducing the pitch joint torques (figure 12(c)). Compared to Controller P + BC (A), Controller P + BC (B) with a smaller degree of controlled conformation helped the robot achieve the same 100% success rates in all cases (figure 12(a)), with reduced up and down oscillations (Movie 1) but increased clearance (figure 12(b)). In contrast, the large degree of controlled conformation in Controller P + BC (C) resulted in larger up and down oscillations than when using P + BC (A) which stalled the shape propagation (Movie 1). This resulted in low success rates in all the cases (figure 12(a)), despite smaller demands on the actuators and the power supply (figure 12(c-d)).

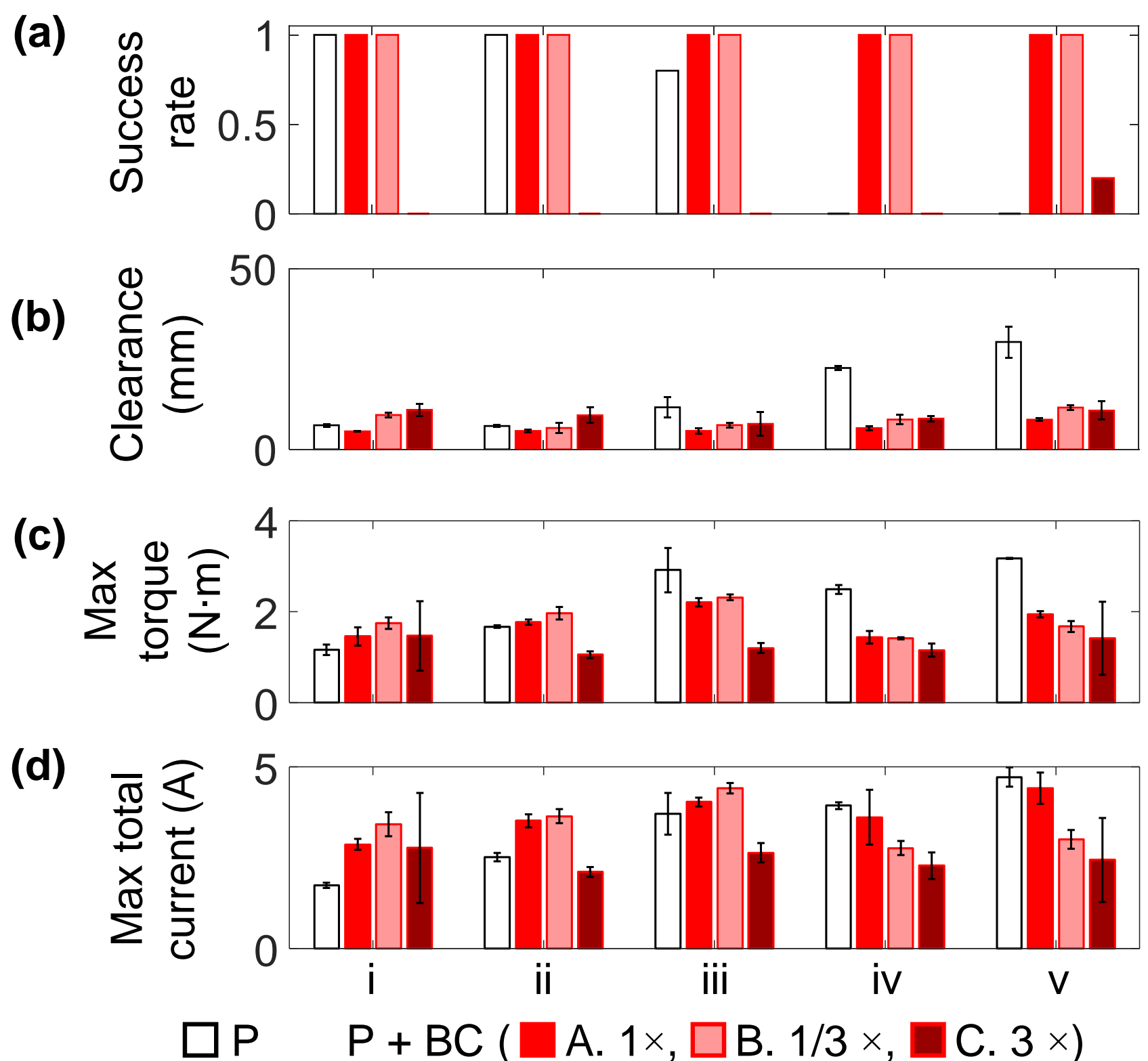

**Figure 12. Comparison of Controllers P and P + BC (A-C) with different degrees of controlled conformation.** **(a)** Success rate. **(b)** Average clearance of all passive wheels. **(c)** Maximal pitch joint torque. **(d)** Maximal total current consumed by pitch joints. Error bars show ± 1 s.d.

To understand the effect of active pushing, we compared Controllers P + HC and P + HC + AP (A-B), which have a zero, small, or large degree of active pushing against the terrain, respectively. When using

Controller P + HC + AP (A) with a small degree of active pushing, the robot lifted more links off the terrain (figure 13(b); Movie 1) to apply the additional active pushing than when using Controller P + HC (section 3.4.1). This resulted in larger demand on the motors and the power supply (figure 13(c, d); section 3.4.2) and more frequent motor stalling (92% vs 0% of the trials) in all the cases, and a lower success rate when the backward load was large (figure 13(a), iii). When the robot used Controller P + HC + AP (B) with a large degree of active pushing, it frequently lost substantial contact with the terrain, sometimes leading to early termination of the trial to prevent collision between the links (Movie 1, v). This resulted in much larger clearance (figure 13(b)) and lower success rates than when using Controllers P + HC and P + HC + AP (A) (figure 13(a)).

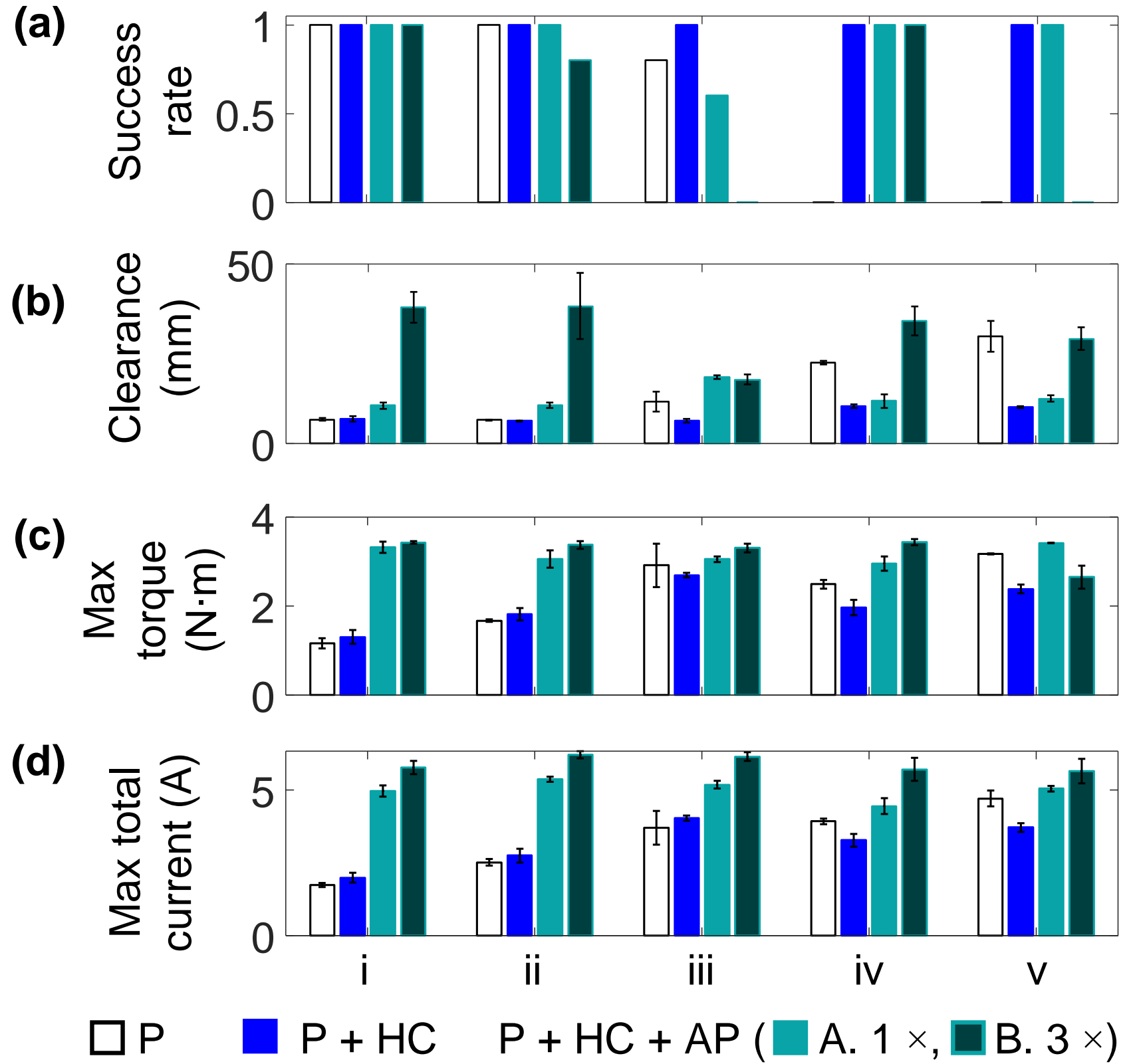

**Figure 13. Comparison of Controllers P, P + HC, and P + HC + AP (A-B) with different degrees of active pushing.** **(a)** Success rate. **(b)** Average clearance of all passive wheels. **(c)** Maximal pitch joint torque. **(d)** Maximal total current consumed by pitch joints. Error bars show ± 1 s.d.

## 4. Discussion

### 4.1. Contribution

Our study advanced the understanding of how snake robots should use vertical bending for propulsion to traverse terrain with large height variation. We compared the distinct bending patterns and body-terrain interaction generated by a robophysical model using different controllers under different perturbations. We discovered that while a robot can generate propulsion using vertical bending using feedforward control, adding contact feedback control can increase the robustness of this strategy against various backward loads or novel terrain geometry. Specifically: (1) On continuous terrain with large bump(s), vertical bending alone allows the robot to generate propulsion to overcome friction and additional backward load. This can be realized by propagating a vertical bending shape posteriorly to push against height variation such as steep downhills if the shape conforms to the terrain. (2) Feedforward shape propagation fails from loss of contact easily under perturbations such as unknown terrain geometry or additional backward load. The loss of contact can lead to loss of propulsion or motor stalling. (3) The robot can use contact feedback control to increase traversal success rate under such perturbations, by sensing the change in contact and modulating bending patterns to: (a) regain contact with steep downhills to maintain propulsion, and (b) avoid long suspending body sections to reduce pitch joint torques. (4) These improvements can be achieved by adding contact feedback-controlled conformation toward the terrain below the body. However, excessive controlled conformation can interrupt shape propagation and reduce propulsion. (5) Adding contact feedback-controlled pushing against the terrain, which is realized by bending the body more concavely against push points behind the body, increases the demands on the actuators and power supply, which increases failures.

Bending the body vertically to push against uneven terrain may be a common strategy for generalist snakes to move in the 3-D world, but this propulsion mechanism has rarely been studied and is still poorly understood [32]. Our study using a robophysical model suggests that snakes likely use contact feedback control to further enhance the robustness of this strategy against unexpected loss of contact. These may inform future investigations of how snakes use tactile sensing to control slithering in the 3-D world. For example, the high success rate of Controller P + HC indicates that the feedback control of the head may play an important role in exploring new environments, leading not only lateral bending [5], [6] but also vertical bending. The high success rates of Controllers P + BC (A-B) indicate similar benefits of whole-body tactile sensing, which may be validated by observing how snakes react to various perturbations with reduced tactile stimulus response after anesthesia [55]. While in this study the robot did not have higher success rates when using whole-body tactile sensing than when using head tactile sensing only, the advantage likely exists when change of contact conditions are not sensible by the head (Movie 3). In addition, all the four feedback controllers used joint angle feedback and Controller P also controls joint angles directly. This suggests that proprioceptive feedback from stretch receptors [37] is likely important for snakes to control vertical bending to push against the environment for propulsion. Finally, considering that the feedback-controlled conformation used in Controllers P + HC and P + BC(A-B) was along the downward direction, snakes may rely on gravity to deform their elastic body to achieve similar conformation without sensory feedback [56], [57].

### 4.2. Comparison between propulsion generation using propagation of a lateral and a vertical bending shape

Despite the similarity in pushing against suitably oriented body surfaces, propulsion generation using lateral and vertical bending has important differences. A snake or a snake robot that propagates a lateral bending shape pushes against push points lateral to the body such as plant stems and rocks. While a snake can push against lateral push points no matter whether they are on both [4]–[6] or only one side [4], [58], [59] of the body, control principles have only been extensively studied when lateral push points are on both sides [2], [33]. In contrast, a snake that propagates a vertical bending shape in the previous studies only pushes against push points below the body, such as uneven terrain (figure 1(b-c)) and horizontal branches (figure 1(a)). This is likely because push points above the body are fewer and harder to push against considering the limited dorsal pitch range of motion of a snake's vertebrae [60], although some snake robots have utilized them under manual control [29], [30].

The difference in the push points results in different environmental forces for a snake or a snake robot to coordinate to move forward. When using lateral bending for propulsion, a snake or a snake robot mainly needs to coordinate forces in the horizontal plane, including contact forces from push points and friction from the ground. For a vertically bending snake or snake robot, it needs to coordinate these forces and body weight in the vertical plane. Body weight always pulls every part of the body toward the terrain and results in upward contact forces. Unlike contact forces that can be modulated by changing bending patterns [2], body weight is constant. Thus, for the common situation in which push points above the body are unavailable, the sum of vertical contact forces is also constant during quasi-static movement because of the force balance. Meanwhile, the torque from the weight of a long suspended body section can overload actuators lifting this section (figure 8(b)). For a snake or snake robot that uses lateral bending strategies, losing contact with terrain surfaces lateral to the body is less catastrophic.

### 4.3. Limitations and future work

Although this study used a robophysical model to understand whether and how a snake or a snake robot can use vertical bending for propulsion with contact feedback control, the performance of the robot can be improved by further engineering the hardware and the controller and by better understanding the neuromechanics of snakes.

First, the hardware of the robot can be upgraded by using high-accuracy sensors and developing more continuous structures. Force sensors with smaller hysteresis like load cells [61] may improve the performance of controllers and enable more accurate analyses of the dynamics (supplementary material 4.1). Torque sensing and control can be improved by using back-drivable motors or introducing torque sensors on the output shaft [45], [62]. However, extra efforts are needed to reduce the sizes of these sensors to fit them inside the snake robot without interfering with body bending. Compared to our robot with discrete sensorized wheels that can be easily caught by terrain asperities and cannot sense forces applied to other parts of the body, a snake has a highly flexible and smooth body [63] with more mechanoreceptors distributed over it [36]. Snake robots with a similarly smooth body [21], [61], [64]–[67] covered by more force sensors [61], [67] are promising in traversing complex environments like snakes do.

In addition, our controllers can be improved by further tuning their parameters. The speed of propagation was selected conservatively to protect the robot and can be increased, which may also reduce the energetic cost (supplementary material 4.2) by increasing locomotion speed. Finer sweeps of the degree of controlled conformation, degree of active pushing, and other control parameters may provide a better understanding of their effects on robot performance and allow further optimization. Using different parameters for different body sections [68] may allow each body section to better adapt to the variable slope of the local surface that it contacts.

Aside from tuning controller parameters, a fruitful next step is to take inspiration from the animals and combine centralized and decentralized control. For a centralized controller like the ones implemented in this study, despite the advantage of coordinating the whole-body motion, the control frequency was significantly limited by the communication between the central controller and the sensors and actuators [69]. Using decentralized control to process detailed sensory information for local feedback control can reduce the amount of information transmitted between the central controller and distributed body components for faster response [12], [69], [70].

Furthermore, although we used uneven terrain and variation of backward loads or terrain geometry to emulate different environmental conditions, the natural environment may pose more complex challenges to vertical bending strategies. For example, it remains to be investigated how to handle dynamic environmental changes such as sudden yielding of loose sand or movement of compliant tree branches. On terrain with limited supporting structures such as tree branches, unconstrained conformation to the terrain may be inefficient and can lead to falling. For terrain with abundant vertical push points, such as rocky beaches, utilizing only some of the push points available with limited body bending may be sufficient [31] and more energetically efficient. In 3-D environments, we need to understand how a snake or a snake robot bends in three dimensions to fully exploit available terrain surfaces for propulsion [31] and coordinate and transition between such 3-D bending at different body sections.

## Acknowledgments

We thank Qihan Xuan, Yifeng Zhang, Henry Astley, Eric Lara, and Divya Ramesh for discussions; Laura Paez for helpful suggestions on servo motor control, and two anonymous reviewers for helpful comments. This work was supported by an Arnold and Mabel Beckman Foundation Beckman Young Investigator Award, a Burroughs Wellcome Fund Career Award at the Scientific Interface, and a Johns Hopkins University Catalyst Award.

**Supplementary Notes**

## 1. Additional details of the robophysical model

### 1.1. Connection of servo motors

Because daisy chaining motors limits the total current supplied to the robot, we connected the two power lines of each servo motor to two cables that directly drew power from a DC power supply (TekPower TP3005DM, Tektronix, Beaverton, OR, USA) at 14 V. The servo motors were connected to the Ubuntu desktop computer via an RS485 bus and a USB communication converter (U2D2, ROBOTIS, Lake Forest, CA, USA). The motors were controlled using the Robot Operating System (ROS Noetic).

### 1.2. Active wheel and IMU

To push the robot up the bump to the initial position before it gains contact with the initial vertical push point (section 2.3), we added an active wheel to the tail of the robot (diameter = 87 mm; figure S1(a), magenta) lifted by the most posterior pitch joint motor and rotated by another servo motor (Dynamixel XM430-W210-R, ROBOTIS, Lake Forest, CA, USA). The motor spinning the active wheel was connected to the computer in the same way as the other motors. However, it was only controlled to rotate at a constant speed of 0.72 rad/s or stay idle and did not provide readings of the present angle or current.

To sense the direction of gravity, we installed an inertial measurement unit (IMU; BNO055 breakout, Adafruit, New York, NY, USA) to the last motor (figure S1). The pitch angle of the IMU $\phi_{IMU}$ was used together with joint angle readings to estimate the pitch angle $\phi_i$ of each link on the computer: $\phi_i = \phi_{IMU} - \pi + \phi_{offset} + \sum_{j=i+1}^{10} \theta_j$ (figure S1(b)).

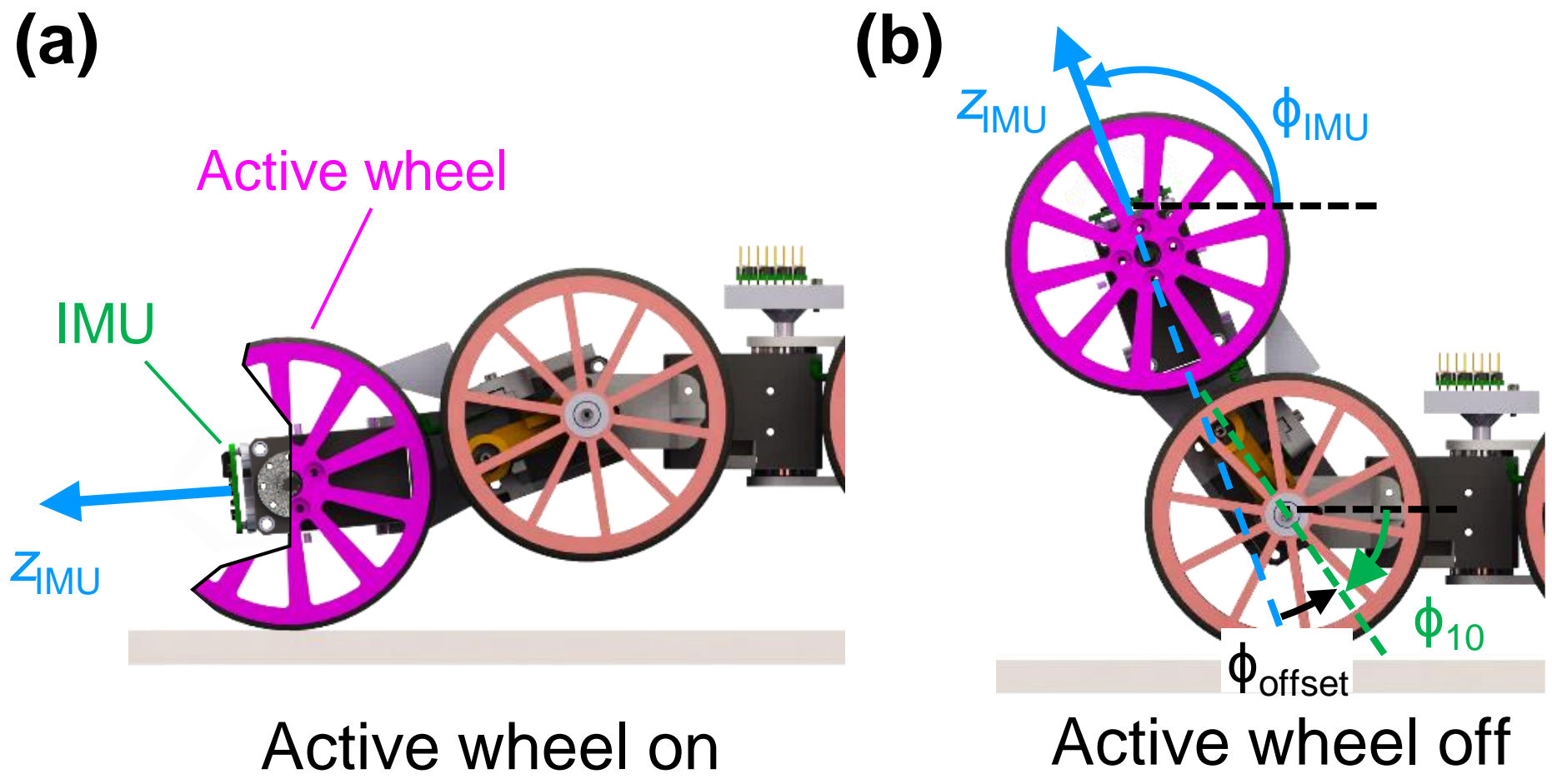

**Figure S1. Structure of active wheel module. (a-b)** Configuration of the active wheel module before (a) and during locomotion experiments (b). Broken out section shows IMU (green), with its $z$-axis (cyan arrow) pointing backward along the centerline of the last pitch joint motor (cyan dashed), which has an angle $\phi_{offset}$ = 15° with the tenth link (green dashed). $\phi_{10}$ and $\phi_{IMU}$ are the pitch angles (positive if clockwise) of link 10 and the IMU, respectively. Magenta and red wheels are active and passive wheels, respectively.

**1.3. Measurement of fore-aft friction coefficient**

To measure the fore-aft friction coefficient of the robot wheels against the terrain covered with rubber sheets, we dragged the robot longitudinally with a constant load and measured its acceleration (figure S2(a)). The robot was initially placed straight on the flat ground covered by the same rubber sheets used in the locomotion experiments. A weight was then connected to the tail of the robot via a string through a pulley system to drag it longitudinally backward. We tracked an ArUco marker attached to the robot at a rate of 60 frame/second and obtained its displacement $x$ as a function of time $t$. We then calculated the acceleration of the robot $a$ by fitting a quadratic function to robot displacement $x(t) = 1/2\ at^2$. The friction coefficient was calculated as $\mu = (m_2 g - (m_1+m_2)a)/(m_1 g)$, where $m_1$ is the mass of the robot, $m_2$ is the mass of the weight, and $g$ = 9.81 m/s$^2$ is the gravitational acceleration. We found that $\mu = 0.14 \pm 0.00$ (mean ± s.d. of three trials).

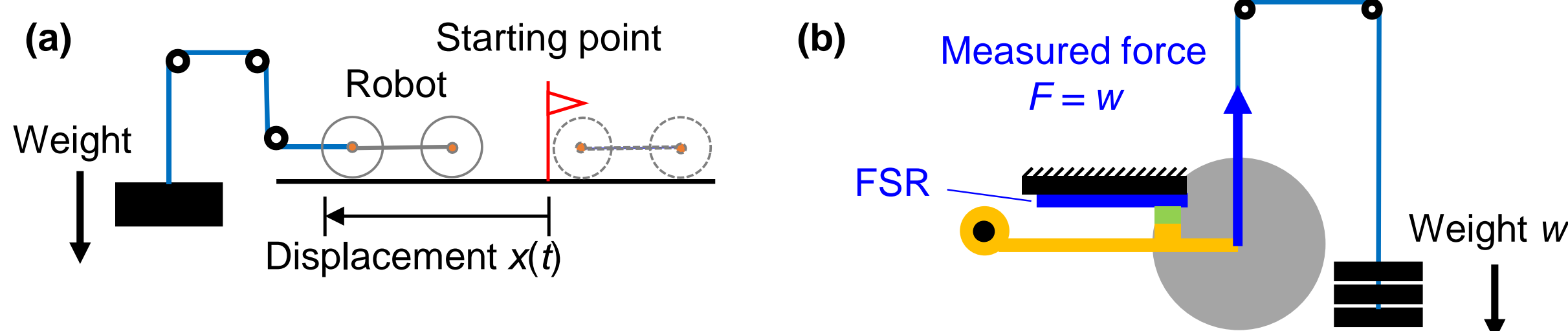

**Figure S2. Experimental setup for measuring friction coefficient and for calibration of force sensing resistor.** **(a)** Setup to measure fore-aft friction coefficient. **(b)** Setup to calibrate force sensing resistor. Gray circle, dark yellow shape, and green square show wheel, wheel arm, and force sensing resistor, respectively. Also see figure 2(b) for detailed structures.

### 1.4. Installation of force sensing resistors

To fixate and protect the sensor, we installed it on a flat surface (figure 2(b), black rectangle) on a 3-D printed mount fixed to a robot link. This surface is parallel to the line segment connecting the two pitch joints on two ends of the link such that the sensor measures the force component normal to the link (figure 2(b)). Each wheel was installed on an arm (figure 2(b), dark yellow) that can freely rotate around a shaft on the link (figure 2(b), black circle). The range of motion of the arm was bounded by the FSR above the arm and a 3-D printed stopper below it. The center of each wheel coincided with the nearest pitch joint when the wheel was pressed against the terrain (figure 2(b)). The center of each wheel could move downward by at most 3 mm with the rotation of the arm about the arm shaft under gravity before reaching the stopper when the wheel was not pressed against the terrain by the body, which guaranteed that the FSR always measured zero force in this situation. A rubber pad (figure 2(b), green) was added to the wheel arm to evenly distribute the exerted force over the force-sensitive area of each FSR and absorb collisional impact. Each FSR and a serially connected resistor $R_M$ = 10 kΩ were supplied with a constant voltage $V_{ref}$ = 3.3 V from the Arduino board to form a measurement circuit (figure S3(a)). Four 4-channel analog-to-digital converters (ADCs) (ADS1015 breakout, SparkFun, Boulder, CO, USA) and four analog input pins on the microcontroller board were used to collect the voltage output $V$ of all the 20 measurement circuits.

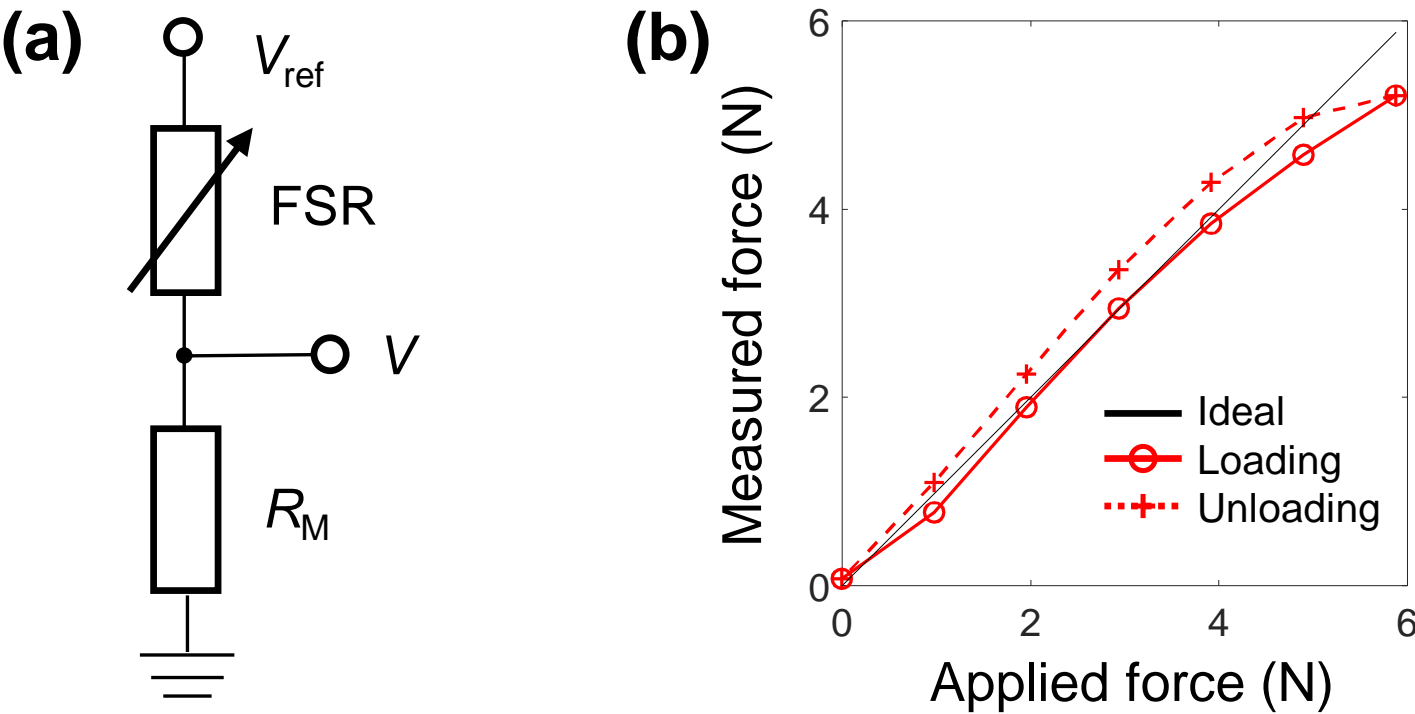

**Figure S3. Circuit of force sensing resistors and example result of calibration. (a)** Measurement circuit of FSR. FSR is serially connected to a constant resistor $R_M$. Circuit is supplied with a constant voltage $V_{ref}$ and outputs a voltage $V$ which is measured by analog-to-digital converters or Arduino. **(b)** Example result of FSR calibration. Measured force using the result of calibration is plotted versus applied force during loading (red solid) and unloading (red dashed) in calibration.

To convert the voltage reading $V$ to the measured force $F$, the resistance of each force sensor $R$ was first calculated using the equation $R = (V_{ref}/V - 1) \cdot R_M$. Because we observed a nearly linear relationship between the force $F$ and the resistance $R$ in the logarithmic scale (figure S4(a)), the measured force $F$ was empirically calculated by $\log F = k_{FSR} \cdot \log \mathrm{R} + \log g + b_{FSR}$, where $k_{FSR}$ and $b_{FSR}$ were fitted in calibration (next section), $g$ is the gravitational acceleration (9.81 m/s$^2$). Each force sensor was calibrated after being installed on the robot (see the next section). During locomotion experiments, if the measured force gave a negative value because of fitting errors in the calibration, it was set to zero. If the measured force was larger than 20 N, the maximum force-sensitive range of the FSR, due to external forces larger than the range, it was set to 20 N.

### 1.5. Calibration of force sensing resistors

To calibrate the force sensing resistors before locomotion experiments, we first fixed the motors directly to an 8020 beam using 3-D printed clamps so that none of the wheels was contacting the ground. Each wheel was pushed against the FSR using slotted weights via a pulley system (figure S2(b)). We increased the force $F$ applied to the wheel from 0 to 5.88 N (loading, figures S3(b) and S4(a), red solid) and then back to 0 (unloading, figures S3(b) and S4(a), red dashed) with an increment of 0.98 N. Resistance of force sensing resistor $R$ corresponding to force $F = w$ was measured 5 seconds after each change of weight, where $w$ is the total mass of slotted weights and $g = 9.81$ m/s$^2$ is the gravitational acceleration. Hereafter, we refer to this cycle as the calibration cycle. The constants $k_{FSR}$ and $b_{FSR}$ were calculated by fitting a line to log $F$ and log $R$ data (figure S4(a), blue). Measurements when $F = 0$ were not used for fitting because the force is too small to actuate the sensor.

**1.6. Characterization of the force sensing resistors after sustained constant load and a sequence of varied loads**

To test whether the force sensing resistors creep or fatigue and change the readings significantly, we performed two calibration cycles before and after applying a specific load through the pulley system (figure S2(b)). We used two types of loads: sustained constant load and a sequence of varied loads. The sustained constant load (figure S4(c)) was applied by hanging a 0.3 kg weight ($F = 2.94$ N) for 30 min. The sequence of varied loads (figure S4(d)) was applied by manually pulling the string downward for 30 seconds. For each case, we performed 3 trials using 3 different force sensing resistors. We then compared a few paired readings corresponding to the same external force applied in the two calibration cycles using paired $t$-tests pooling all 9 trials. The paired readings included the readings when the force $F$ equaled 0.98 N during the loading process (first non-zero step) of the two calibration cycles, the readings for all the steps in the two calibration cycles, and the reading when the force $F$ equaled 0.98 N during the unloading process (last non-zero step) of the two calibration cycles.

We found that the readings before and after a sustained constant load were statistically the same either for all the steps inside a calibration cycle pooled together ($t(116) = 1.205$, $P = 0.23$, paired $t$-test) or for the first non-zero step in the cycle ($t(8) = 1.075$, $P = 0.31$, paired $t$-test). Therefore, the creep after long time loading can be neglected during our locomotion experiments.

Readings in all the steps inside a calibration cycle pooled together were 0.09 N larger before the sequence of varied loads than after the impact ($t(116) = 2.918$, $P < 0.005$, paired $t$-test). However, the reading of the last non-zero step after the sequence of varied loads was the same as before the impact ($t(8)$

= – 0.067, $P$ = 0.95, paired $t$-test). This implies that the fatigue caused by the sequence of varied loads is small and can recover during the 5-min interval between trials.

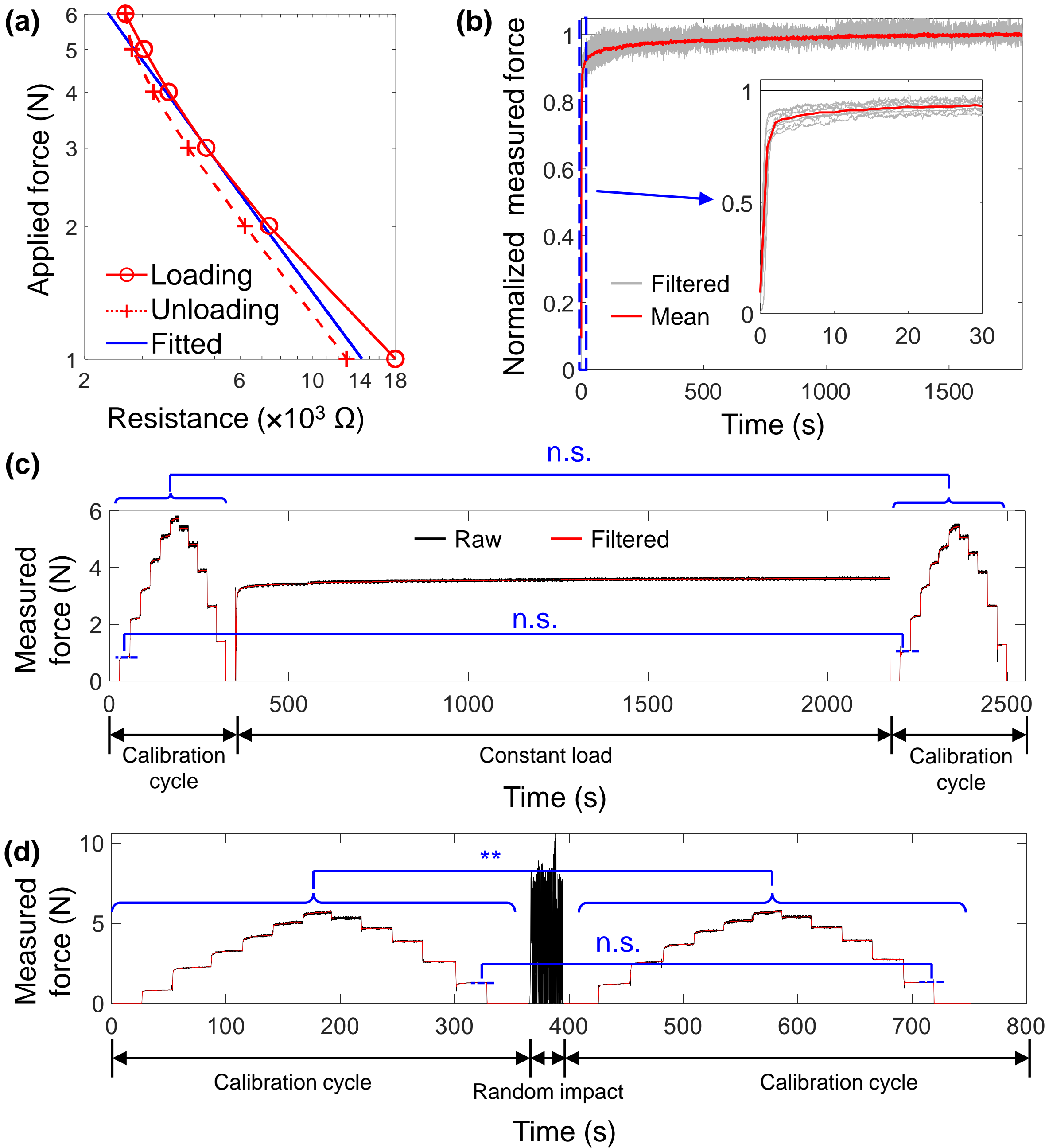

**Figure S4. Characteristics of force sensing resistors. (a)** Force as a function of resistance during calibration, plotted in log scales. Red solid and dashed curves show measured resistance as a function of forces applied during loading and unloading in calibration, respectively. Blue line shows linear fit log $F$ =

$k_{FSR} \cdot \log R + \log g + b_{FSR}$ of data. **(b)** Force as a function of time under a load of 2.94 N over 30 minutes. Gray curves shows filtered results from all 9 trials and red curve shows average across trials. **(c)** Force as a function of time during a test to characterize effects of a sustained load. **(d)** Force as a function of time during a test to characterize effects of a sequence of varied loads. Brackets and asterisks represent statistically significant differences between readings before and after applying a specific load (**$P < 0.005$, paired $t$-test).

## 2. Experimental protocol

### 2.1. Construction of terrain with large height variation

We used the downhill of a bump with a cylindrical upper surface to provide vertical push points (figure 6(a)). The 0.49 m long, 0.12 m high bump was fixed to the ground and made by gluing together laser-cut 6.35 mm thick wooden sheets (McMaster-Carr, Elmhurst, IL, USA). The bump and the ground were covered by a rubber sheet (EPDM 60A 1.6 mm thick rubber sheet, Rubber-Cal, Fountain Valley, CA, USA) to increase traction. The additional bump for variation of terrain geometry in case (v) was made using the same method but has a different size, which is 0.13 m high and 0.25 m long.

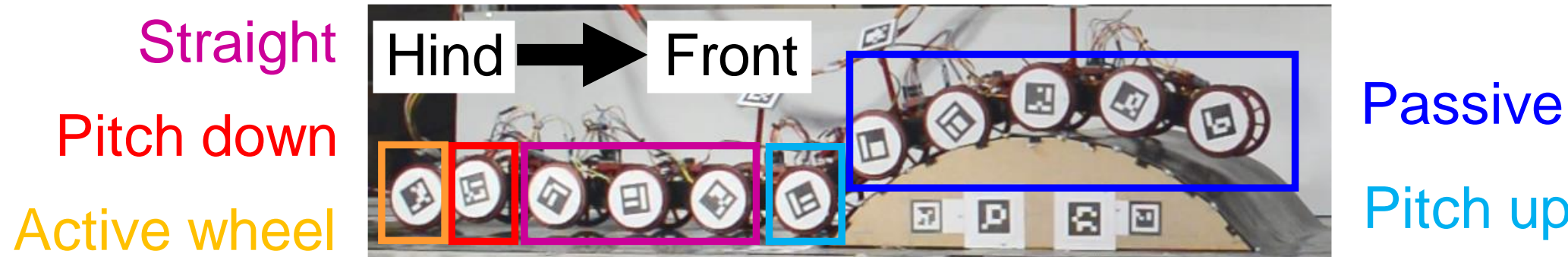

**Figure S5. Initial climbing to position the robot into the initial position of locomotion experiments using an active wheel.** Torques are turned off for pitch joints above the main bump (blue box) and on for the other pitch joints. As the robot is propelled forward by the active wheel (orange box) and hits the bump using the most anterior passive wheel behind the bump (cyan box), the shape and torque patterns are propagated down the body by one link length (all boxes shift down body by one link length).

### 2.2. Climbing process before a trial

Before each trial, the robot was initially placed horizontal and straight behind the bump. The robot first pushed its active wheel against the ground (figure 2(b), right; figure S5, orange) by bending the last pitch joint (figure S5, red) down by 20°. Meanwhile, all the links between the last pitch joint and the pitch joint closest to the bump were held straight (figure S5, purple) to increase the contact force exerted by the active wheel. As the spinning active wheel pushed the robot forward, the joint closest to the bump pitched up (figure S5, cyan) to climb up the bump, while the pitch joints anterior to it had its torque turned off (figure S5, blue) to passively conform to the bump under gravity. When the most anterior passive wheel behind the bump (cyan box) started pushing against the bump, the angle and torque conditions of all the pitch joints were propagated down the robot by one link. This process repeated until the third pair of passive wheels from the tail contacted the bump (figure 6(a)). Then, the active wheel stopped spinning and was lifted by the last joint pitched up by 75° to reduce friction (figure 2(b), right). Then, all the other pitch joint torques were disabled to let the robot conform to the terrain under gravity. This ensured that the initial shape of the robot was the same for all the robot controllers within each of the five treatments. The initial joint angles were recorded, and motor torques were enabled. This climbing process was prescribed and the propagation was manually triggered by visual checking of the contact between the wheel in the cyan box and the bump.

### 2.3. Performing a trial in different cases

A trial started when the vertical bending controller started. A trial was ended if the robot traversed the entire trackway, moved backward and detached the main bump (due to the backward load), stopped bending after a controller finished propagation all the way through the body, bent into a shape that potentially led to flipping over or self-collision between different body components, was no longer able to move forward after motor stalling, or got stuck at the same position for over 15 seconds and after two cycles of periodic behaviors (such as up and down oscillations of pitch joints). If any structural part in the robot (motor connectors, wheels, wheel arms, etc.) broke from impact with the terrain, the trial was rejected and re-collected after replacing the broken part.

In cases (ii) and (iii) with a small and large backward load, respectively, we pulled the tail of the robot backward by hanging a weight using a string through a pulley system (figure 6(a)). The pulley system was placed 1 m away from the robot tail at the beginning of a trial, so that the direction of the backward force remains horizontal (with a variation of around 3°) despite the up-and-down displacement of the robot tail. Weights of 150 and 300 g were used to generate a small and large load (1.5 and 2.9 N), respectively. The load was applied to the robot in each trial after the initial shape was recorded and torque was enabled but before the controllers started working.

In case (iv) with an imperfect initial shape, we added a 0.3 m long acrylic plate below the robot initially at the same location for all trials (figure 6(b), left). Stones were placed on two sides of the plate to hold the plate in place. The robot climbed over the plate and passively conformed to it under gravity before the initial joint angles were recorded. The plate and the stones were then removed after the torque was enabled (figure 6(b), right) but before a trial began.

In case (v) with an unknown bump, we added an additional half-cylindrical bump in front of the main bump (figure 6(c)). The rear edge of the additional bump was 0.29 m in front of the front edge of the main bump. The additional bump was also fixed to the ground and covered with the same rubber sheet to ensure the same surface condition.

During the locomotion experiments, the cables powering and controlling the robot were bundled together and routed by a wheel carriage above the robot. This was to ensure that the cables always stayed vertically above the tail under manual control to minimize their effects on propulsion. An operator monitored the cables and used a long stick to push or pull the wheel carriage along a horizontal frame (McMaster-Carr, Elmhurst, IL, USA).

**2.4. Data collection**

To track the positions of the wheels, we attached an ArUco marker [49] to each wheel on the right side of the robot. To track the terrain geometry for calculation of the clearance, we attached two ArUco markers to the main bump, one ArUco marker to the additional bump, and five ArUco markers to the horizontal ground. The other markers were attached to the top of the robot and to the terrain but not used during the analyses. Four synchronized cameras (N-5A100-Gm/CXP-6-1.0, Adimec, Eindhoven, The Netherlands) recorded locomotion at 60 frame/s with a resolution of 2592 × 2048 pixels. To correct lens distortion, we calculated the distortion parameters of each lens using a checkerboard and the MATLAB Camera Calibrator application. Snapshots and videos recorded by the four cameras were undistorted using the MATLAB Computer Vision Toolbox when the snapshots and the videos were used for calibration and tracking. To calibrate the cameras for 3-D reconstruction, we placed a 61 × 66 cm calibration object made of DUPLO bricks (The Lego Group, Billund, Denmark) with BEEtag markers [50] in the imaging area each time the cameras were booted and adjusted before locomotion experiments. We tracked the markers and calculated the 2-D coordinates of the centers of the markers in each camera view. The coordinates were used to calculate intrinsic and extrinsic camera parameters necessary for 3-D reconstruction using direct linear transform [51]. The ArUco markers were tracked in each camera view and reconstructed for their 3-D positions and orientations using the calibrated camera parameters. Bad tracking results of the ArUco markers attached to the wheels were rejected by checking whether the angle between the marker plane and the vertical plane (figure 6(a), $x$-$z$ plane) was larger than 30°. We used this threshold to reject false tracking results that typically rotated the maker by more than 45° while accounting for the small rotation of the marker planes resulting from manufacturing inaccuracy. Then, the missing 3-D positions and orientations of each ArUco marker were filled temporally using a linear interpolation of the twist coordinates [52].

All the forces, IMU orientation, present motor angle and current, and motor goal angle and current data were recorded in ROS together with their timestamps. To synchronize the ROS-recorded data with the camera tracking data, the active wheel was turned on, left spinning for more than 1 second, and turned off before and after each trial. We then linearly interpolated both the kinematics data and the ROS-recorded data between the moments when the active wheels started spinning in these two procedures with a time interval of 1/60 s for synchronization with the video frames.

For real-time monitoring and recording of the entire locomotion experiments, a digital camera (D3200, Nikon, Tokyo, Japan) was used to record the locomotion experiments (including the climbing process and the entire trial) from the side view at 25 frame/s with a resolution of 1920 × 1080 pixels.

## 3. Data analyses

### 3.1. Calculation of the performance metrics

#### 3.1.1. Success rate

The success rate $P_{\text{Success}}$ is calculated as: $P_{\text{Success}} = n_{\text{Success}}/\ n_{\text{Total}},$ where $n_{\text{Total}} = 5$ is the number of trials when the robot using one controller in one case and $n_{\text{Success}}$ is the number of successful trials among them.

#### 3.1.2. Clearance

To reconstruct the 3-D terrain profile for calculation of clearance, we first obtained the geometry of the bumps and the ground, then used the tracked ArUco markers to locate the terrain relative to the cameras in each trial. To obtain the terrain geometry, we digitized 3-D positions of 20 and 24 vertices of the main and the additional bumps, respectively, and the corner points of all the ArUco markers attached to the terrain in all four camera views at one moment using DLTdv8 [51]. The ground was then represented by a plane (figure 6(a), *x*-*y* plane) fitted to all the digitized points on the ground. Each bump was represented by a polyhedron using the digitized vertices. If no markers were tracked because of occlusion or poor lighting, additional corner points of the markers were manually digitized using DLTdv8.

We then calculated the clearance of each wheel in the vertical plane (figure 6(a), *x*-*z* plane) in each video frame and averaged them spatiotemporally for each trial. The projection of the terrain in this vertical plane contained the line segments from the projection of flat and those from the projection of top surfaces of the bumps. Clearance was calculated as the closest distance between the wheel center and the terrain [53] deducted by one wheel radius (43.5 mm). When calculating the closest distance, the point on the terrain closest to each wheel center was recorded as a reference point, which was also considered as the contact points if the wheel contacts the terrain. The closet distance was then the distance between the wheel center and the reference point. If multiple contact points were present for one wheel, only one was recorded as the reference point. This is inaccurate for wheels contacting both the bump and the ground, but its effect on the results was minor because such moments were rare from visual examination of side-view recordings. In addition, using either contact point as the reference point will return almost identical result of the closest distance. We also noticed that the wood sheets used to construct the bumps were sometimes deformed from large normal contact forces, but the effect was small (0.6 mm for every 3 N of normal force, which is the average vertical force exerted by one wheel) compared to the total values (> 4 mm on average for all the passive wheels, figure 7(b)).

After obtaining the clearance for each wheel in each video frame, we first averaged it spatiotemporally across all the wheels and all the video frames for each trial, then calculated the mean and standard deviation across all five trials for each controller in each case.

#### 3.1.3. Maximal torque

The maximal torque $\tau_{max}$ in one trial is calculated by finding the maximal torque generated by any pitch joint motor (excluding the one lifting the active wheel) in any video frame. We used the interpolated current data that was synchronized with video frames (supplementary material section 2.4) in this step. The torque $\tau$ was calculated from the current $I$ using the assumed linear relationship $\tau = k_\tau I$, where the torque constant $k_\tau = 1.78\ \mathrm{N \cdot m \cdot A^{-1}}$ [43] (section 2.2.3). We then calculated the mean and standard deviation of the maximal torque across all five trials for each controller in each case.

#### 3.1.4. Maximal total current

We first summed up the current consumed by all the pitch joint motors (excluding the one lifting the active wheel) in one video frame, using the interpolated current data that was synchronized with video frames (supplementary material 2.4). Then, we obtained the maximal total current among all the video frames in this trial and calculated the mean and standard deviation across all five trials for each controller in each case.

### 3.2. Calculation of the terrain reaction force

Each force sensing resistor measured the component of the terrain reaction force normal to one link. The total reaction force $F_r$ was zero if the measured normal force $F$ was zero. To calculate the total reaction force when the wheel was contacting the terrain (i.e., when the measured force was larger than zero), we needed to obtain the direction of the terrain reaction force. We first obtained the direction of normal force (figure 2(c), purple vector), which pointed from the contact point (figure 2(c), purple point) to the tracked center of the wheel. Friction (figure 2(c), green dashed) was then calculated using the friction coefficient $\mu = 0.14$, which was perpendicular to the normal force and opposite to the velocity of the center of this wheel relative to the contact point. We set friction to be zero if the calculated velocity was smaller than 2 mm/s (i.e., with little relative motion to the terrain and likely producing unpredictable static friction). The total terrain reaction force (figure 2(c), red) was then the vector sum of normal force and friction. If the angle between the calculated total terrain reaction force (figure 2(c), red) and the measured normal force (figure 2(c), blue) $\theta_{R,M}$ was larger than 85°, the force data on this wheel at this moment was. This is because for this near-singular configuration, a small measurement error of $\theta_{R,M}$ or the measured force $F$ will result in an excessively large error in the value of total reaction force $F_r = F/\cos(\theta_{R,M})$. Terrain reaction forces were averaged temporally using a moving window size of 0.11s.

We also obtained the relationship between horizontal contact force (propulsion) and vertical contact force (figure S6(a)). When a wheel is pushing against a surface with slope $\beta$ (negative for downhills), the horizontal contact force $F_{r,x} = -N\sin\beta - \mu N\cos\beta$, the vertical contact force $F_{r,z} = N\cos\beta - \mu N\sin\beta$, thus the ratio between them $r = F_{r,x}/F_{r,z} = (-\sin\beta - \mu\cos\beta)/(\cos\beta - \mu\sin\beta)$. This ratio is independent of the normal force $N$. In addition, the ratio monotonically decrease with $\beta$ when $\beta$ is between -90° and 0° because the derivative $dr/d\beta = -1 - ((\sin\beta + \mu\cos\beta)/(\cos\beta - \mu\sin\beta))^2$ is always negative regardless of friction coefficient $\mu$ (figure S6(b)).

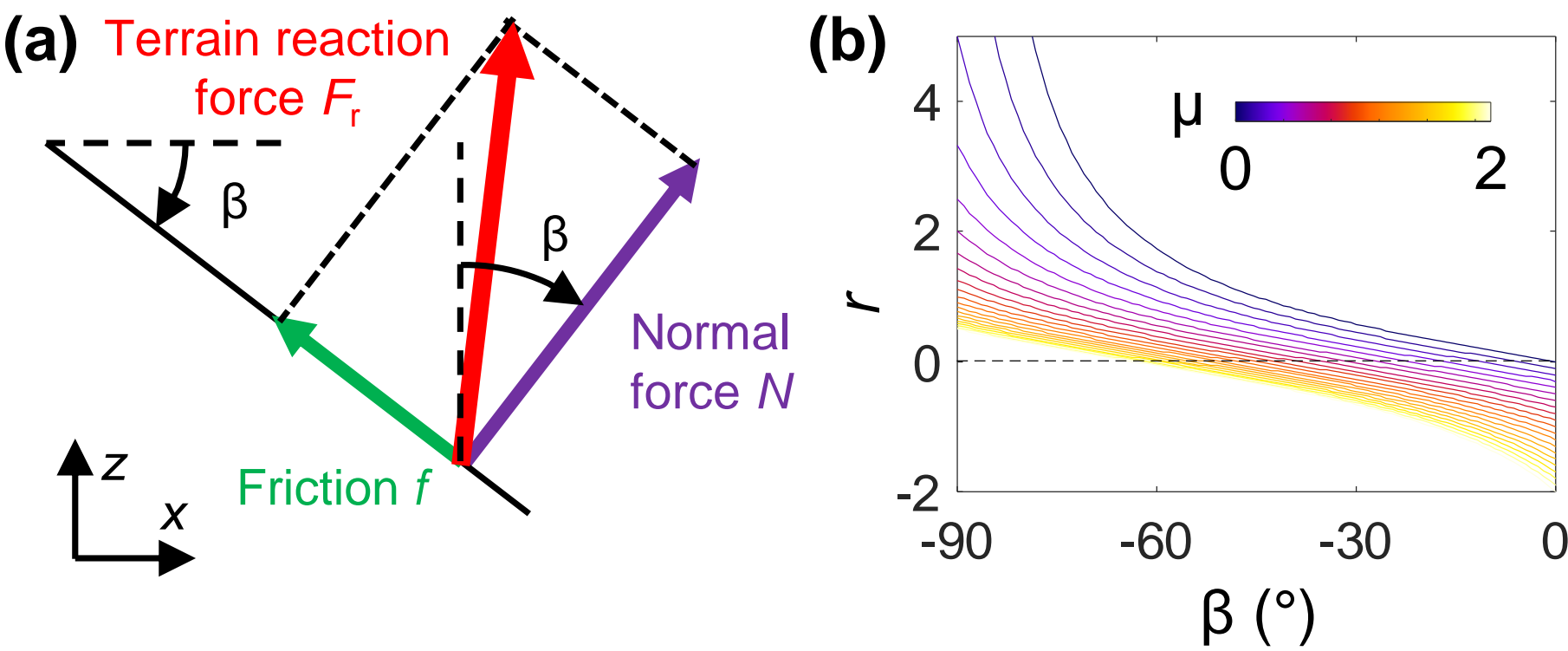

**Figure S6. Relationship between horizontal and vertical contact force. (a)** Schematics of contact forces. Black solid line shows terrain surface in contact and β is its slope (negative for downhills). **(b)** Ratio *r* between horizontal (positive along +*x*-axis) and vertical (positive along +*z*-axis) component of terrain reaction force as a function of slope β. μ is friction coefficient between wheels and the terrain.

### 3.3. Analyses on modulation of propulsion by controlling contact conditions using body bending

To understand how body bending patterns affected body-terrain contact, we performed statistical tests to test the dependence of contact force on the number of suspended links or on the pitch joint torque, two controllable parameters of the bending pattern. Specifically, we tested the linear correlation between the number of links $n$ in the longest suspended body section and the sum of vertical contact forces $F_{r, z}$ on its two ends (figure 9), and between the maximum pitch joint torque τ and the vertical contact force $F_{r, z, 1}$ on the wheel closest to this joint (figure 10). For each pair of data, we performed a linear regression between them after pooling all data from the video frames from all the trials.

We tried two different criteria to determine a suspended body section in each video frame and performed the linear regression twice using each of them: (1) a body section was considered suspended if the measured contact forces on all the wheels inside it (not including the two wheels on its two ends) equaled zero, and (2) a body section was considered suspended if the measured torque of every pitch joint inside it (not including the two joints on its two ends) was negative (clockwise). For example, in figure 9(a), the suspended link with 5 links can be identified by checking that the contact forces on the 4 wheels inside it all equal to zero or the torques of the 4 pitch joints inside it are all negative. After identifying the suspended body sections, the number of links inside the longest one was obtained for each video frame.

We obtained the maximum pitch joint torque τ and the vertical contact force $F_{r, z, 1}$ on the wheel closest to this joint for each video frame. The maximum pitch joint torque τ was defined as the measured torque or the pitch joint that was generating the largest torque in this video frame.

We also derived the relationship between the maximum pitch joint torque τ and the vertical contact force $F_{r, z, 1}$ on the wheel installed at the same position in an idea situation (figure 10(a)). We assumed the robot in this situation is contacting the horizontal ground with only three wheels, is moving quasi-statically, and is horizontal and straight. The force balance along the vertical direction and the torque balance about the middle pitch joint whose wheel is contacting the ground are as follows:

$$\begin{cases} F_{r,z,0} + F_{r,z,1} + F_{r,z,2} = (n_f + n_b)w \\ -F_{r,z,2} \cdot n_b L - \tau + F_{r,z,0} \cdot n_f L = 0 \end{cases}$$

where $w$, $L$ are the weight and length of each link, $n_f$, $n_b$ are the number of suspended links on two sides of the middle wheel, and $F_{r,z,1}$ is the vertical contact force on the middle wheel. Using these balance functions, we derived the positive torque τ as a function of the vertical contact force $F_{r,z,1}$: $\tau = (n_f n_b/(n_f + n_b))LF_{r,z,1} - ((n_f + n_b)/2)mgL$.

To understand how body-terrain contact affected propulsion generation, we also performed statistical tests to test whether the slope of the contact surface of interest was steeper when the robot was moving (head velocity larger than 3 mm/s) and when it was stuck in place (head velocity no larger than 3 mm/s). The head velocity was defined as the velocity of the first pair of wheels along the *x*-axis (figure 6(a)) calculated after smoothing the *x*-position of the wheels temporally using a window average filter (window size = 0.3 s). We performed the test for two types of surfaces, the downhill where the robot experienced the largest vertical contact force $F_{r,z}$ and the steepest downhill that the robot was contacting. To obtain the slopes of these two surfaces in each video frame, we looped through all the passive wheels and identified the one with the largest vertical contact force $F_{r,z}$ and the one whose reaction force $F_r$ was positive and the slope β of the surface it was contacting was the smallest. To test whether the slope differed between the two types of video frames, we performed analysis of variance after pooling all the video frames in each condition.

## 4. Additional data analyses and results

### 4.1. Total propulsion

To calculate the propulsion generated by pushing against the terrain, we summed the horizontal components (positive along the +*x*-axis in figure 6(a)) of terrain reaction forces acting on the wheels (figure 2(c), red). The average propulsion within a trial was then calculated by averaging the sum of propulsion from all the wheels across all the video frames in each trial. We then calculated the mean and standard deviation (s.d.) across all the trials in each case for each controller.

We found that for our robot that moved quasi-statically most of the time, the total propulsion may not be a good metric for evaluation of its performance. After deducting the additional backward load applied (1.5 or 2.9 N in case (ii) or (iii), respectively), the average net horizontal force experienced by the wheels in different cases (figure S7(a)) was in the range of kinetic friction, which was approximately 0 ± 4.2 N, and did not show a clear relationship with the success rates. This variation around 0 was likely because we could not estimate the static friction without measuring it using dedicated force sensors. The inaccuracy of force sensing resistors (figure S3(b)) also likely contributed to this variation.

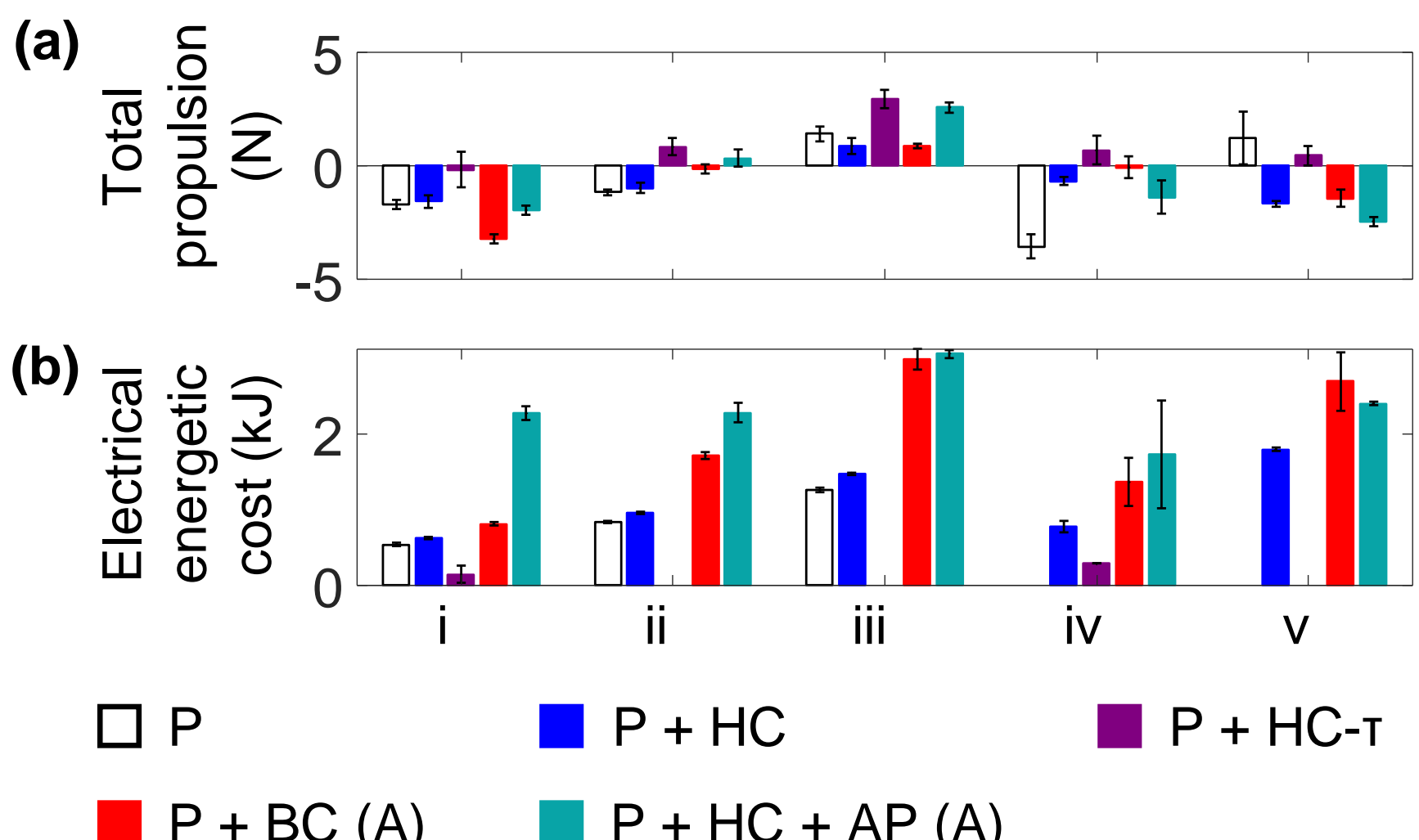

**Figure S7. Additional measurement of the robot locomotion.** **(a)** Total propulsion. **(b)** Electrical energetic cost of pitch joints in successful traversal. Error bars show ± 1 s.d.

#### 4.2. Electrical energetic cost

To evaluate the total electrical energetic cost by the robot to traverse the bump(s) using vertical bending, we first calculated the electrical energy consumed by each pitch joint motor in each video frame by multiplying its present current, voltage, and video frame duration. We then summed the total energetic cost by all the pitch motors in all the video frames in each successful trial and calculated their mean and standard deviation across all the trials in each case for each controller.

Despite higher success rates and better adaptation to the terrain, the robot consumed more electrical energy to complete a successful traversal when using the feedback Controllers P + HC, P + BC (A), and P + HC + AP (A) than when using the feedforward Controller P (figure S7(b)). For Controllers P + HC and P + BC (A), the energy also increased with additional backward loads in cases (ii-iii) or when additional adaptation to the terrain was needed because of the variation in terrain geometry in cases (iv) and (v) compared to in case (i). The electrical energy (32~3171 J) was substantially higher than the mechanical work needed to drag the robot forward on flat ground covered by the same rubber sheet (6 J for the same displacement of 1.4 m on the flat ground without additional load), which implies a need of improving locomotion efficiency for our robot using vertical bending for propulsion.

## 5. List of movies

**Movie 1. Representative trials of the robot using different controllers in different cases.** One representative trial is shown for the robot using each controller in each case. Failed trials are presented whenever they are available to demonstrate the failure modes. However, we show a successful trial instead of a failed one for the robot traversing the track under a small backward load (ii) when using Controller P + HC + AP (B) to demonstrate the possibility to complete traversal despite motor stalling. We also show a successful trial instead of a failed one for the robot traversing the track under zero backward load (i) when using Controller P + HC-τ to demonstrate the rare success when using this controller.

**Movie 2. Effects of the gain of pitch joint torque in Controller P + HC-τ.** A smaller gain $K_P$ results in smaller up and down oscillations of the pitch joints but also lower success rates. A larger gain results in higher success rates under no loads, but faster oscillations that damages the robot in most of the trials and the same zero success rate when there is a medium load.

**Movie 3. Ability of the robot using the basic controllers to conform to terrain variation behind the head.** Similar to in case (iv) with an imperfect initial shape, an acrylic plate (yellow dashed) was added before the initial shape was recorded and removed before the controllers started. Only Controller P + BC

(A) succeeded in traversing the terrain (blue).

**Movie 4. Ability of the robot using contact feedback controllers to traverse a large bump with little initial contact with the downhill.** When only the head was initially contacting the vertical push point (the downhill), the robot was able to traverse the bump with Controller P + HC, P + BC (B), and P + HC + AP (A). The robot did not traverse the bump using this position and shape when using the other controllers in our preliminary tests.